\documentclass[epj]{svjour}
\usepackage{graphics}
\usepackage{graphicx}
\usepackage{hyperref}
\usepackage{epsfig}
\usepackage{amsmath}
\usepackage{amssymb}
\usepackage{bm}
\usepackage{cite}
\usepackage[scriptsize]{subfigure}

\def\abs#1{|#1|}
\begin{document}
\title{Thermoelectric transport through a finite-$U$ quantum dot side-coupled to Majorana bound state}

\author{Chol Won Ri \thanks{e-mail: cholwon\_ri@163.com} \and Kum Hyok Jong \and Song Jin Im \and Hak Chol Pak}

\institute{Department of Physics, Kim Il Sung University, Ryongnam Dong, Taesong District, Pyongyang, Democratic People's Republic of Korea}

\date{Received: 08 September 2018 / Revised version: date}
\abstract{
We study the thermoelectric transport through a single-level quantum dot (QD) coupled to two normal metallic leads and side-coupled to Majorana bound state (MBS).
The Coulomb interaction in QD is considered.
To investigate only the influence of MBS on thermoelectric transport, we focus on the relatively high temperature region ($T\gg T_K$), where Kondo effect does not appear.
The electric and thermal conductance and thermopower as a function of gate voltage  (i.e. QD level) are completely different whether the coupling between MBSs is zero or not.
When the coupling between MBSs is finite, all thermoelectric characteristics are similar to the transport without MBS.
However, for zero MBSs' coupling, the electric and thermal conductance peaks are reduced by 3/4.
Especially, in the case of QD without MBS, the sign of thermopower changes three times, however, in the case of QD strongly side-coupled to ideal and isolated MBS, the sign of thermopower changes 9 or 5 times.
It can be used for detecting of the signature of MBS.
It has actual possibilities when the nanowire is long enough and pure without any defects.
\PACS{
      {74.25.Fy}{Transport properties} \and
      {73.63.Kv}{Quantum dots} \and
      {74.45.+c}{Proximity effects; Andreev reflection; SN and SNS junctions}   \and
      {74.78.Na}{Mesoscopic and nanoscale systems}
     } % end of PACS codes
} %end of abstract
\maketitle

%%%%%%%%%%%%%%%%%%%%%%%%%%%%%%%%%%%%%%%

\section{Introduction}\label{intro}

Majorana fermion is a particle that is its own antiparticle, which was predicted by Ettore Majorana \cite{Ettore_Nuovo} in the early years of relativistic quantum mechanics.
Majorana fermion has been attracting lots of attention in condensed matter physics, due to its exotic nature, distinct with Dirac fermion, and its characteristics providing the fault-tolerant topological quantum computing \cite{Elliott_RMP87, Nayak_RMP80, Hyart_PRB88, Landau_PRL116, Aasen_PRX6, Plugge_NJP19, Karzig_PRB95, Guessi_PRB96, Manousakis_PRB95}.
It is one of the open problems to find the Majorana fermion as an elementary particle in high energy physics, while it was suggested that it can exist as a quasi-particle in condensed matter physics, hence experimental efforts are dedicated to prove it \cite{Mourik_SCI336, Das_NP8, Deng_NL12, Deng_SCI354, Devillard_PRB96, Danon_PRB96}.
Unpaired Majorana fermions can be localized in certain range when the band structure of one-dimensional p-wave superconductor is topologically non-trivial (see e.g. Ref. \cite{Bernevig_Book, Shen_Book}).
For example, Kitaev \cite{Kitaev_USP44} showed that unpaired and localized Majorana fermions (Majorana Bound States - MBSs) can be appeared in two ends of 1D p-wave superconductor which is topologically non-trivial.
It can be achieved by attaching the semiconducting nanowire (InSb, InAs, etc.) with strong spin-orbit coupling into proximity with conventional s-wave superconductors (Al, Nb, etc.) and subjecting the external magnetic field \cite{Elliott_RMP87,Bernevig_Book, Shen_Book,Leijnse_SST27}.
For topologically non-trivial, the Zeeman splitting should be satisfied that $\abs{E_z}>\sqrt{\mu^2+\Delta^2}$ (here $\Delta$ is superconducting gap and $\mu$ is the chemical potential of the wire).

Since Majorana fermion is not a real particle, but a quasi-particle, it can be detected by using some indirect effect like transport property.
In particular, it can be regarded as one of the effective methods for detecting MBS to use the quantum dot (QD).
To study MBSs in the ends of 1D p-wave superconductor (topological superconductor-TSC), there are lots of researches about electron transport through several structures such as normal metallic lead (NL)/QD/TSC \cite{Leijnse_PRB84, Golub_PRL107}, NL/QD/TSC/QD/NL \cite{Lu_PRB86}, QD side-coupled to TSC \cite{Liu_PRB84, Cao_PRB86, Lee_PRB87, Vernek_PRB89}, T-shaped multiple QDs \cite{Huo_EPJB89, Napitu_EPJB88}, and so on.
In the case of spinless QD side-coupled to TSC, the zero-bias voltage peak of conductance is reduced by half than original unitary limit due to the combination with QD and MBS \cite{Liu_PRB84} and the zero frequency part of shot noise is increased due to MBS \cite{Cao_PRB86}.
In the Kondo regime, however, the QD-MBS coupling makes the unitary-limit value of the linear conductance 3/4 \cite {Lee_PRB87}.

Thermoelectric transport is also one of the best routes to detect the MBS \cite{Leijnse_NJP16, Lopez_PRB89, Khim_EPJB88, Shapiro_PRB95, Ramos-Andrade_PRB94, Hou_PRB88}.
Leijnse \cite{Leijnse_NJP16} showed that NL/QD/MBS structure can be used for detecting MBS by measuring the gate-dependent Seebeck coefficient.
In spinless QD side-coupled to MBS, the sign of the thermopower is changed and the both of the electrical and thermal conductance are reduced by half by being attached MBS to QD \cite{Lopez_PRB89}.
The thermoelectric transport through the Kondo QD side-coupled to MBS was also studied \cite{Khim_EPJB88}.

Now there is no doubt for the existence of MBS.
The problem is how the characteristics of thermoelectric transport through QD attached to MBS are in detail.
Furthermore, the characteristics of thermoelectric transport through QD side-coupled to MBS will be changed much differently by the existence of MBS and Coulomb interaction.
For example, in the absence of MBS the sign of the thermopower as a function of gate-voltage is changed once in spinless QD \cite{Lopez_PRB89}, however, it changes three times when Coulomb interaction in QD is considered \cite{Zimbovskaya_JCP140, Swirkowicz_PRB80}.
So we can predict that change of the sign of the thermopower will become more complicated and interested due to the presence of MBS in such that system.
In practice, it is also important to consider the QD with Coulomb interaction, instead of spinless QD, in the transport through the QD attached to MBS (more details will be discussed in Sect. \ref{sec:model}). 

In this paper we study a problem --- the thermoelectric transport through a single-level QD side-coupled to MBS, where Coulomb interaction in QD is considered.
The paper is organized as follows.
Sect. \ref{sec:model} presents the model together with the formulas used to study thermoelectric characteristics and details some technical aspects related to the calculation of the QD Green function.
Sect. \ref{sec:discussion} and Sect. \ref{sec:conclusions} present our results and conclusive discussion.

%%%%%%%%%%%%%%%%%%%%%%%%%%%%%%%%%%

\section{Model and Methods} \label{sec:model}

\begin{figure}[h]
\begin{center}
  \includegraphics[width=6.0cm]{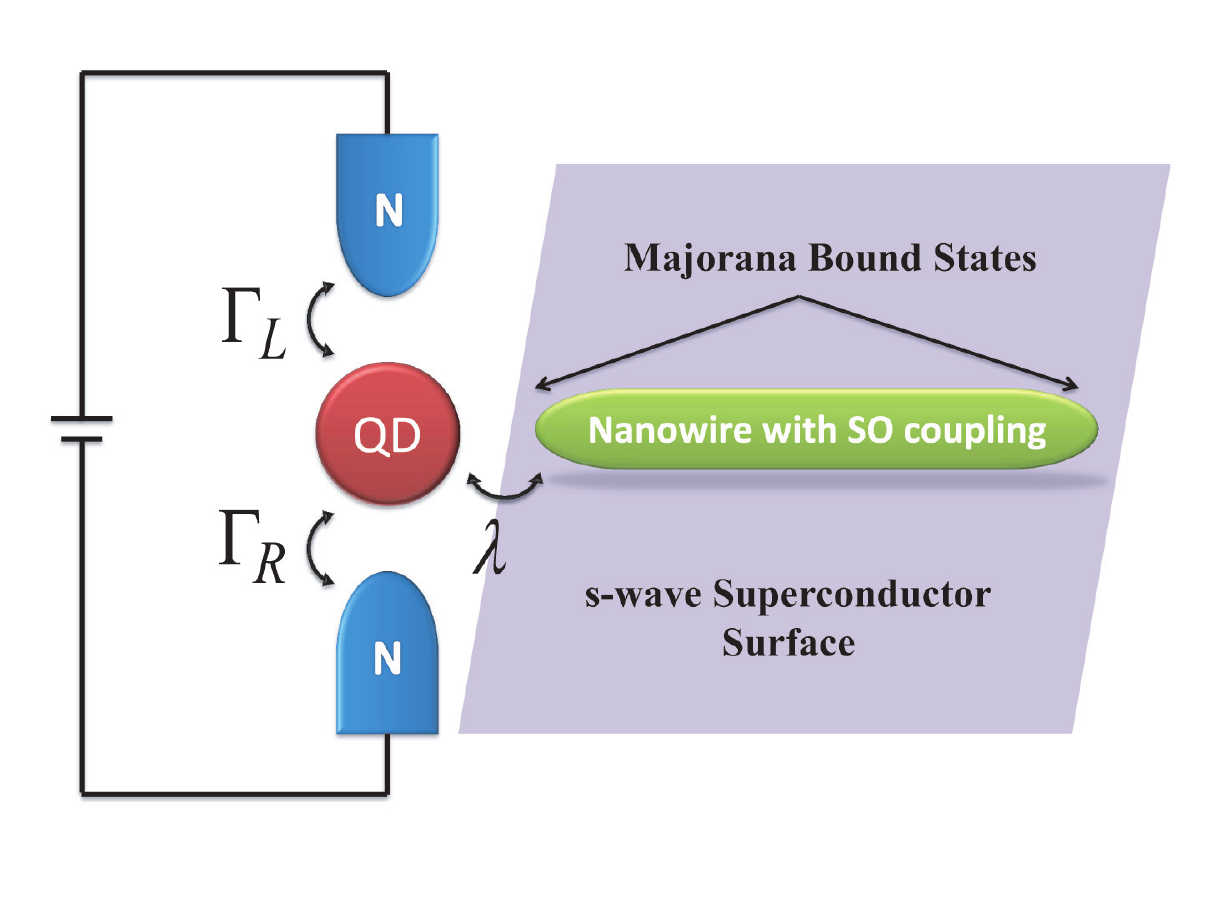}
\end{center}
\caption{ 
The QD coupled to two metallic leads and side-coupled to MBS \cite{Liu_PRB84}. 
Due to the proximity effect with s-wave superconductor and the strong magnetic field applied whole system, the nanowire with spin-orbit coupling becomes one-dimensional topological superconductor (1D TSC) phase that isolated MBSs appear at two ends of the wire.}
\label{fig:model}
\end{figure}

We consider a single-level QD coupled to two metallic leads and side-coupled to an 1D topological superconductor, suggested by D. E. Liu {\em et. al} \cite{Liu_PRB84}.
The isolated Majorana fermion zero modes appear at two ends of nanowire with strong Rashba spin-orbit interaction due to the proximity-induced s-wave superconductor and the strong magnetic field applied whole system (see Fig. \ref{fig:model}).

Many previous studies (see e.g. Ref. \cite{Liu_PRB84, Cao_PRB86}) assumed that the spin degrees of freedom in QD can be ignored, i.e. QD can be regarded as spinless (or spin-polarized) QD due to the presence of strong magnetic field.
However, the Zeeman splitting by the external magnetic field is not so large in many experiments.
Let us take the recent experiment \cite{Deng_SCI354} as an example, which was studied the electron transport in the N/QD/TSC structure.
There InAs nanowire was covered by epitaxial Al for almost region of nanowire and QD was made by very small bared InAs region at the end of nanowire.
At that time, the interested parameters were given as following: the Coulomb interaction in QD is $U\sim6$meV, the effective superconducting gap is $\Delta^*\sim0.2$meV, the effective Land\'e factor is $g^*\sim$4, the critical magnetic field of s-wave superconductor is $B_{\rm{C}}\sim2.2$T, the threshold of magnetic field for making the nanowire topologically non-trivial is $B_{\rm{C}, \rm{topo}}\sim1$T and the maximum magnetic field in experiment is $B\sim2$T. 
So, the maximum value of Zeeman splitting for maximum field $B\sim2$T is $E_z=g\mu_B B\sim0.5$meV. 
It is satisfied the condition for topologically non-trivial nanowire, $\abs{E_z}>\sqrt{\mu^2+\Delta^2}$, because the superconducting gap of nanowire is $\Delta^*\sim0.2$meV and the chemical potential of nanowire is gate-controlled.
Therefore the Zeeman splitting in QD is rather smaller than Coulomb interaction $U$ and we should consider but two spin component QD containing the Coulomb interaction between spin-opposite electrons rather than spinless QD. 

The whole system can be described by the Hamiltonian given by:

\begin{equation} \label{eq:Hamiltonian}
H=H_{NL}+H_{QD}+H_{NL\textrm{-}QD}+H_{MBS}+H_{MBS\textrm{-}QD}.
\end{equation}
Here $H_{NL}=\sum_{k\beta\sigma}\epsilon_{k\beta\sigma}c^{\dagger}_{k\beta\sigma}c_{k\beta\sigma}$ describes the non-interacting left $(\beta=L)$ and right $(\beta=R)$ normal metallic leads,  $\epsilon_{k\beta\sigma}$ is the single-electron energy in the $\beta$-th lead for wave vector $k$ and electron spin $\sigma=(\uparrow, \downarrow)$ and $c^{\dagger}_{k\beta\sigma}$($c_{k\beta\sigma}$) denotes the corresponding creation(annihilation) operator. 
The second term $H_{QD}=\sum_{\sigma}\epsilon_{\sigma}d^{\dagger}_{\sigma}d_{\sigma}+Ud^{\dagger}_{\uparrow}d_{\uparrow}d^{\dagger}_{\downarrow}d_{\downarrow}$ describes the single-level QD and here $\epsilon_{\sigma}$ is the electron energy in QD for spin $\sigma$, whereas $d^{\dagger}_{\sigma}(d_{\sigma})$ is corresponding creation(annihilation) operator. 
In the presence of the external magnetic field, the energy level in QD $\epsilon_d$ is splitting by $\epsilon_{\sigma}=\epsilon_d+\sigma E_z$, where $E_z=g\mu_B B$ is Zeeman splitting. The third term, $H_{NL\textrm{-}QD}=\sum_{k\beta\sigma}(T_{k\beta\sigma} c^{\dagger}_{k\beta\sigma} d_{\sigma}+T^*_{k\beta\sigma}d^{\dagger}_{\sigma} c_{k\beta\sigma})$, describes the tunnelling between normal leads and QD, where $T_{k\beta\sigma}$ is the component of tunnelling matrix coupling between $\beta$-th lead and QD for electron energy $\epsilon_{k\beta\sigma}$. 
The next term $H_{MBS}=i \epsilon_M \eta_1 \eta_2$ describes the MBSs at ends of 1D TSC nanowire, where $\eta_1$ and $\eta_2$ are Majorana fermion zero mode operators being satisfied $\eta_i=\eta^{\dagger}_i$, $\eta^2_i=1$ and $\{\eta_i,\eta_j\}=2\delta_{ij}$.
And $\epsilon_M \sim e^{-L/\xi}$ is coupling between MBS $\eta_1$ and $\eta_2$, where $L$ is the length of the wire and $\xi$ is superconducting coherence length.
The last term $H_{MBS\textrm{-}QD}=\sum_{\sigma}(\lambda_{\sigma} d_{\sigma}-\lambda^*_{\sigma} d^{\dagger}_{\sigma})\eta_1$ describes the coupling between QD and MBS, where  $\lambda_{\sigma}$ describes the coupling between QD electron with energy $\epsilon_{\sigma}$ and nearby MBS $\eta_1$.

By using the nonequilibrium Green function technique \cite{Haug_Book}, in the presence of the bias voltage and the difference of temperature between two normal leads, the electric current $I$ and the thermal current $Q$ from left to right lead can be written as following \cite{Costi_PRB81, Liu_PRB81, Trocha_PRB85}.

\begin{eqnarray}\label{eq:current}
\left(\begin{array}{cc}I \\ Q \end{array}\right)=-\frac{1}{\hslash}\int d E \left(\begin{array}{cc} -e\\ E-\mu_L \end{array}\right)\frac{\Gamma_L\Gamma_R}{\Gamma_L+\Gamma_R}\textrm{DOS}(E)\times\nonumber
\\
\times [f_L(E)-f_R(E)],
\end{eqnarray}
where $\Gamma_{\beta}=2\pi\sum_{k}\abs{T_{k\beta\sigma}}^2 \delta(E-\epsilon_{k\beta\sigma})$ describes the contribution to the half-width of QD level due to tunnelling through the $\beta$-th lead, $f_{\beta}(E)=1/\{\exp[(E-\mu_{\beta})/k_B T]+1\}$ is the Fermi-Dirac distribution in the $\beta$-th lead, $\textrm{DOS}(E)=\sum_{\sigma}i [G^r_{\sigma}(E)-G^a_{\sigma}(E)]/2\pi$ is the density of state (DOS) in QD and $G^{\eta}_{\sigma}(E)=\langle\langle d_{\sigma};d_{\sigma}^{\dagger}\rangle\rangle^{\eta}_{E} \quad (\eta=r, a)$ are the Fourier transforms of the retarded and advanced Green function of QD electron, respectively.
In the limit of linear response and in the presence of small chemical potential difference $\delta\mu=\mu_L-\mu_R$ and small temperature gradient $\delta T=T_L-T_R$, electric current $I$ and thermal current $Q$ obey following linear equations \cite{Costi_PRB81, Liu_PRB81, Trocha_PRB85, Mahan_Book}: 
\begin{equation}\label{eq:onsager}
\left(\begin{array}{cc} I \\ Q \end{array}\right)=\left(\begin{array}{cc} L_{11}\quad L_{12}\\ L_{21}\quad L_{22}\end{array}\right)
\left(\begin{array}{cc} -\frac{\delta\mu}{T} \\-\frac{\delta T}{T^2}\end{array}\right),
\end{equation}
where $L_{ij}$ $(i,j=1,2)$ are the kinetic coefficients, being $L_{11}=I_{0}$, $L_{12}=L_{21}=I_1$, $L_{22}=I_2$, while 
\begin{equation}\label{eq:def_I}
I_n=-\frac{T}{\hslash}\int d E\frac{\Gamma_L\Gamma_R}{\Gamma_L+\Gamma_R}\textrm{DOS}(E)(E-\mu)^n\left(\frac{\partial f}{\partial E}\right),
\end{equation}
where $T=T_L=T_R$, $\mu=\mu_L=\mu_R$.
The characteristics of thermoelectric transport, the electric conductance $G$, the thermal conductance $\kappa$ and the thermopower (Seebeck coefficient) $S$ can be determined as following \cite{Costi_PRB81, Liu_PRB81, Trocha_PRB85, Mahan_Book}:
\begin{eqnarray}
\begin{aligned}
G=&\frac{e^2}{T}L_{11} \\
\kappa=&\frac{1}{T^2}\left(L_{22}-\frac{L^2_{12}}{L_{11}}\right) \\
S=&-\frac{1}{eT}\frac{L_{12}}{L_{11}}.\label{eq:charact_th}
\end{aligned}
\end{eqnarray}

To determine these characteristics one should calculate the retarded Green function of QD $G^r_{\sigma}(E)=\langle\langle d_{\sigma};d^{\dagger}_{\sigma}\rangle\rangle^r_E$.
It can be calculate by using the equation of motion (EOM) method \cite{Haug_Book} in framework of nonequilibrium Green function techniques.

It is very difficult to calculate the retarded Green function by EOM method due to the presence of MBS and Coulomb interaction in QD, therefore it is very convenient to introduce the 4-component Nambu spinor formalism as following:
\begin{eqnarray}
\begin{aligned}
\bar{\gamma}&=(d_{\uparrow},\quad d^{\dagger}_{\downarrow},\quad d_{\downarrow},\quad d^{\dagger}_{\uparrow})^{T},\\
\bar{\psi}_{k\beta}&=(c_{k\beta\uparrow},\quad c^{\dagger}_{k\beta\downarrow},\quad c_{k\beta\downarrow},\quad c^{\dagger}_{k\beta\uparrow})^T,\\
\bar{\chi}&=(\eta_1, \quad \eta_2, \quad \eta_2, \quad \eta_1)^{T},\label{eq:Nambu_def}
\end{aligned}
\end{eqnarray}
where $\bar{\gamma}$, $\bar{\psi}$, $\bar{\chi}$ describe the QD, normal metal lead (NL) and MBS, respectively. 
At first, the EOM for QD Green function $\bm{G}(E)=\langle\langle\bar{\gamma};\bar{\gamma}^{\dagger}\rangle\rangle_E$ is 
\begin{equation}\label{eq:eom_1st}
(\bm{E}-\bm{\epsilon}_D)\bm{G}(E)=\bm{I}+\sum_{k\beta}\bm{T}^{\dagger}_{k\beta}\bm{K}_{k\beta}(E)-\bm{\Lambda}^{\dagger}\bm{L}(E)+\bm{U}\bm{G}^{(2)}(E),
\end{equation}
where $\bm{K}_{k\beta}(E)=\langle\langle\bar{\psi}_{k\beta};\bar{\gamma}^{\dagger}\rangle\rangle_{E}$ $\big( \bm{L}(E)=\langle\langle\bar{\chi};\bar{\gamma}^{\dagger}\rangle\rangle_{E}\big)$ is NL (MBS)-QD Green function, $\bm{G}^{(2)}(E)=\langle\langle\bar{\gamma}^{(2)};\bar{\gamma}^{\dagger}\rangle\rangle_E$ is 2nd-order QD Green function, $\bar{\gamma}^{(2)}=(d_{\uparrow}n_{\downarrow}, d^{\dagger}_{\downarrow}n_{\uparrow}, d_{\downarrow}n_{\uparrow}, d^{\dagger}n_{\downarrow})^{T}$ is 2nd-order QD spinor, $\bm{E}=E\bm{I}$ and $\bm{I}$ is $4\times4$ identity. And $\bm{\epsilon}_D=diag(\epsilon_{\uparrow}, -\epsilon_{\downarrow}, \epsilon_{\downarrow},-\epsilon_{\uparrow})$, $\bm{T}_{k\beta}=diag(T_{k\beta\uparrow}, -T^*_{k\beta\downarrow}, T_{k\beta\downarrow},-T^*_{k\beta\uparrow})$, $\bm{U}=diag(U,-U,U,-U)$ are the matrices of QD energy, NL-QD coupling, Coulomb interaction, respectively, and $\bm{\Lambda}$ is the matrix of MBS-QD coupling, defined as 
\begin{equation}
\bm{\Lambda}=\frac{1}{2}\left(\begin{array}{cccc}\lambda_{\uparrow} & -\lambda^*_{\downarrow} & \lambda_{\downarrow} & -\lambda^*_{\uparrow}\\0 &  0 & 0 & 0\\0 & 0 & 0 & 0\\ \lambda_{\uparrow} & -\lambda^*_{\downarrow} & \lambda_{\downarrow} & -\lambda^*_{\uparrow}\end{array}\right).\nonumber
\end{equation}
In Eq. (\ref{eq:eom_1st}) the EOM for NL(MBS)-QD Green function $\bm{K}_{k\beta}(E)$ $\big(\bm{L}(E)\big)$ is respectively,
\begin{eqnarray}
(\bm{E}-\bm{\epsilon}_{k\beta})\bm{K}_{k\beta}(E)&=\bm{T}_{k\beta}\bm{G}(E) \\
(\bm{E}-\bm{\epsilon}_{M})\bm{L}(E)&=4\bm{\Lambda}\bm{G}(E),\label{eq:eom_N-QD}
\end{eqnarray}
where $\bm{\epsilon}_{k\beta}=diag(\epsilon_{k\beta\uparrow}, -\epsilon_{k\beta\downarrow}, \epsilon_{k\beta\downarrow}, -\epsilon_{k\beta\uparrow})$ is the matrix of NL energy and $\bm{\epsilon}_M$ is the matrix of coupling between two MBSs, defined as
\begin{equation}
\bm{\epsilon}_{M}=2 i \left(\begin{array}{cccc} 0 & 0 & \epsilon_M & 0\\0 & 0 & 0 & -\epsilon_M \\-\epsilon_M & 0 & 0 & 0\\ 0 & \epsilon_M & 0 & 0\end{array}\right).\nonumber
\end{equation}
The EOM for 2nd-order QD Green function $\bm{G}^{(2)}(E)=\langle\langle\bar{\gamma}^{(2)};\bar{\gamma}^{\dagger}\rangle\rangle_E$ is more complicated.
It has been contained NL(MBS)-QD 3rd-order Green function, such as $\langle\langle c^{\dagger}_{k\beta\uparrow}d_{\uparrow}d^{\dagger}_{\downarrow};d^{\dagger}_{\uparrow}\rangle\rangle_E$ $\big(\langle\langle\eta_1 d_{\uparrow}d^{\dagger}_{\downarrow};d^{\dagger}_{\uparrow}\rangle\rangle_E\big)$.

At this stage, we apply the Hartree-Fock approximation \cite{Haug_Book} (or Hubbard I approximation \cite{Hubbard_PRSLA276}) to decouple the higher-order Green functions $\big($e.g. $\langle\langle c^{\dagger}_{k\beta\uparrow}d_{\uparrow}d^{\dagger}_{\downarrow};d^{\dagger}_{\uparrow}\rangle\rangle_E\approx\langle d_{\uparrow}d^{\dagger}_{\downarrow}\rangle\langle\langle c^{\dagger}_{k\beta\uparrow};d^{\dagger}_{\uparrow}\rangle\rangle_E\big)$.
Of course, this approximation ignores some correlations (quantum fluctuations) that appear at very low temperatures, and therefore can not account for important phenomena such as Kondo effect.
However, if the temperature is much higher than the Kondo temperature, such correlations are very small, so in this case the Hubbard I approximation may be applied.
As we discuss below (see in Sect. \ref{sec:discussion}), the combination with MBS produces small and sharp peak (MBS peak) in the density of state near $E = 0$, whereas Kondo resonance related to the Kondo effect also appears near $E = 0$.
The purpose for this paper is the influence for MBS to the thermoelectric transport, so we focus on the relatively high temperature region ($T\gg {{T}_{K}}$), where only the MBS peak appears and the Kondo peak does not appear.
In this decoupling approximation the EOM for the 2nd-order QD Green function is 
\begin{align}\label{eq:eom_2nd}
\begin{split}
(\bm{E}-\bm{\epsilon}_D-\bm{U})\bm{G}^{(2)}(E) = \langle\tilde{\bm{n}}\rangle&+\langle\tilde{\bm{n}}\rangle\sum_{k\beta}\bm{T}^{\dagger}_{k\beta}\bm{K}_{k\beta}(E) \\
&-\langle\tilde{\bm{n}}\rangle\bm{\Lambda}^{\dagger}\bm{L}(E),
\end{split}
\end{align}
where $\tilde{\bm{n}}$ is the matrix made of elements of number operator matrix $\bm{n}=\bar{\gamma}^{\dagger}\otimes\bar{\gamma}$, defined as
\begin{equation}\label{eq:def_tilde_n}
\tilde{\bm{n}}=\left(\begin{array}{cccc} n_{\downarrow} & d_{\downarrow}d_{\uparrow} & d_{\uparrow}d^{\dagger}_{\downarrow} & 0 \\d^{\dagger}_{\downarrow}d^{\dagger}_{\uparrow} & n_{\uparrow} & 0 & d_{\uparrow}d^{\dagger}_{\downarrow}\\d_{\downarrow}d^{\dagger}_{\uparrow} & 0 & n_{\uparrow} & d_{\uparrow}d_{\downarrow}\\ 0 & d_{\downarrow}d^{\dagger}_{\uparrow} & d^{\dagger}_{\uparrow}d^{\dagger}_{\downarrow} & n_{\downarrow}\end{array}\right).
\end{equation}

The series of equation (\ref{eq:eom_1st})-(\ref{eq:eom_2nd}) is closed, therefore, we can get the QD Green function to solve it: 
\begin{align}
\begin{split}
&\bm{G}(E)=[(\bm{E}-\bm{\epsilon}_D-\bm{U})(\bm{E}-\bm{\epsilon}_D)-
\\
&-(\bm{E}-\bm{\epsilon}_D-\bm{U}+\bm{U}\langle\tilde{\bm{n}}\rangle)\bm{\Sigma}(E)]^{-1}\times [\bm{E}-\bm{\epsilon}_D-\bm{U}+\bm{U}\langle\tilde{\bm{n}}\rangle],
\label{eq:sol_Green}
\end{split}
\end{align}
where $\bm{\Sigma}=\bm{\Sigma}_{NL}+\bm{\Sigma}_{MBS}$ is the self-energy, while $\bm{\Sigma}_{NL}(E)=\sum_{k\beta}\bm{T}^{\dagger}_{k\beta}(\bm{E}-\bm{\epsilon}_{k\beta})^{-1}\bm{T}_{k\beta}$ and $\bm{\Sigma}_{MBS}(E)=4\bm{\Lambda}^{\dagger}(\bm{E}-\bm{\epsilon}_M)^{-1}\bm{\Lambda}$ are the self-energy due to NL-QD and MBS-QD coupling, respectively. The retarded and advanced Green function can be calculated as $\bm{G}^{r/a}(E)=\bm{G}(E\pm i 0^+)$.
The retarded Green function $G^r_{\sigma}(E)=\langle\langle d_{\sigma};d^{\dagger}_{\sigma}\rangle\rangle^r_E$ is the $(1,1)$ and $(3,3)$ element of retarded Green function matrix $\bm{G}^r(E)$.
In order to determine the retarded Green function matrix (\ref{eq:sol_Green}), we should calculate the matrix $\langle\tilde{\bm{n}}\rangle$ (\ref{eq:def_tilde_n}) and for it, the average particle number matrix $\langle\bm{n}\rangle$, which is defined as:
\begin{equation}{\label{eq:mean_particle}}
\langle\bm{n}\rangle=\int d E \frac{\Gamma_L f_L(E)+\Gamma_R f_R(E)} {\Gamma_L+\Gamma_R}\bm{DOS}(E),
\end{equation} 
where $\bm{DOS}(E)$ is the matrix of DOS in QD:
\begin{equation}{\label{eq:def_DOS}}
\bm{DOS}(E)=\frac{i}{2\pi}(\bm{G}^r(E)-\bm{G}^a(E))
\end{equation}
and it's (1,1) and (3,3) elements are the local density of state of up- and down-spin electron in QD, respectively.
The average particle number matrix $\langle\bm{n}\rangle$ (\ref{eq:mean_particle}) and the retarded Green function matrix $\bm{G}^r(E)$ (\ref{eq:sol_Green}) should be calculated self-consistently. 

Note that the Hartree-Fock approximation for calculating the Green function is so lower that the result does not reflect the effects appeared at very low temperature, like Kondo effect.
As a matter of fact, in order to study the Kondo effect, we should use the higher order of approximation.

\section{Result and Discussion}\label{sec:discussion}

For the simplicity we suppose that two metal leads are coupled to QD symmetrically, i.e. $\Gamma_L=\Gamma_R$ and set the chemical potential of lead as the reference of energy, i.e. $\mu=0$. 

According to recent experiment \cite{Deng_SCI354}, we set the parameters for numerical calculation as following.
By supposing the strong coupling between QD and metal leads, we set $\Gamma=\Gamma_L+\Gamma_R\sim 1$meV, and also set the Coulomb interaction in QD, $U \sim 10\Gamma$, the QD-MBS coupling, $\lambda\sim0.5\Gamma$, the coupling between MBSs, $\epsilon_M\sim0.5\Gamma$ and the Zeeman splitting by external magnetic field, $E_z\sim0.4\Gamma$.
The bandwidth of metal leads is about $D = 50\Gamma$, hence all integrations are carried out in the region of $-D\sim D$.
As mentioned in Sect. \ref{sec:model}, the temperature of the system should be much higher than the Kondo temperature in order to ignore the Kondo correlations, but the system contains s-wave superconductor, so the temperature must be lower than superconducting transition temperature.
The Kondo temperature \cite{Hewson_Book, Svilans_PRL121}  ${k_B}{T_K}=\tfrac{1}{2}\sqrt{\Gamma U}\exp \left[ \pi \epsilon_d\left(\epsilon_d+U \right)/\left( \Gamma U \right) \right]$ in the case of $\lambda=0$, $U=10\Gamma$ and $\epsilon_d=-U/2$ is approximately equal to $6.14\times 10^{-4}\Gamma$.
From this consideration, we set the temperature of the system about 1K ($k_B T\sim 0.1\Gamma$).
Note that the Kondo resonance is destroyed on a temperature scale of order $20T_K$ \cite{Dutta_ASCNano19}.
Throughout this paper, $\Gamma$ is chosen as a unit in our calculations.

%\subsection{Density of state}\label{subsec:DOS}
\begin{figure}
	\begin{center}
		\subfigure[][]{\includegraphics[width=4.7cm]{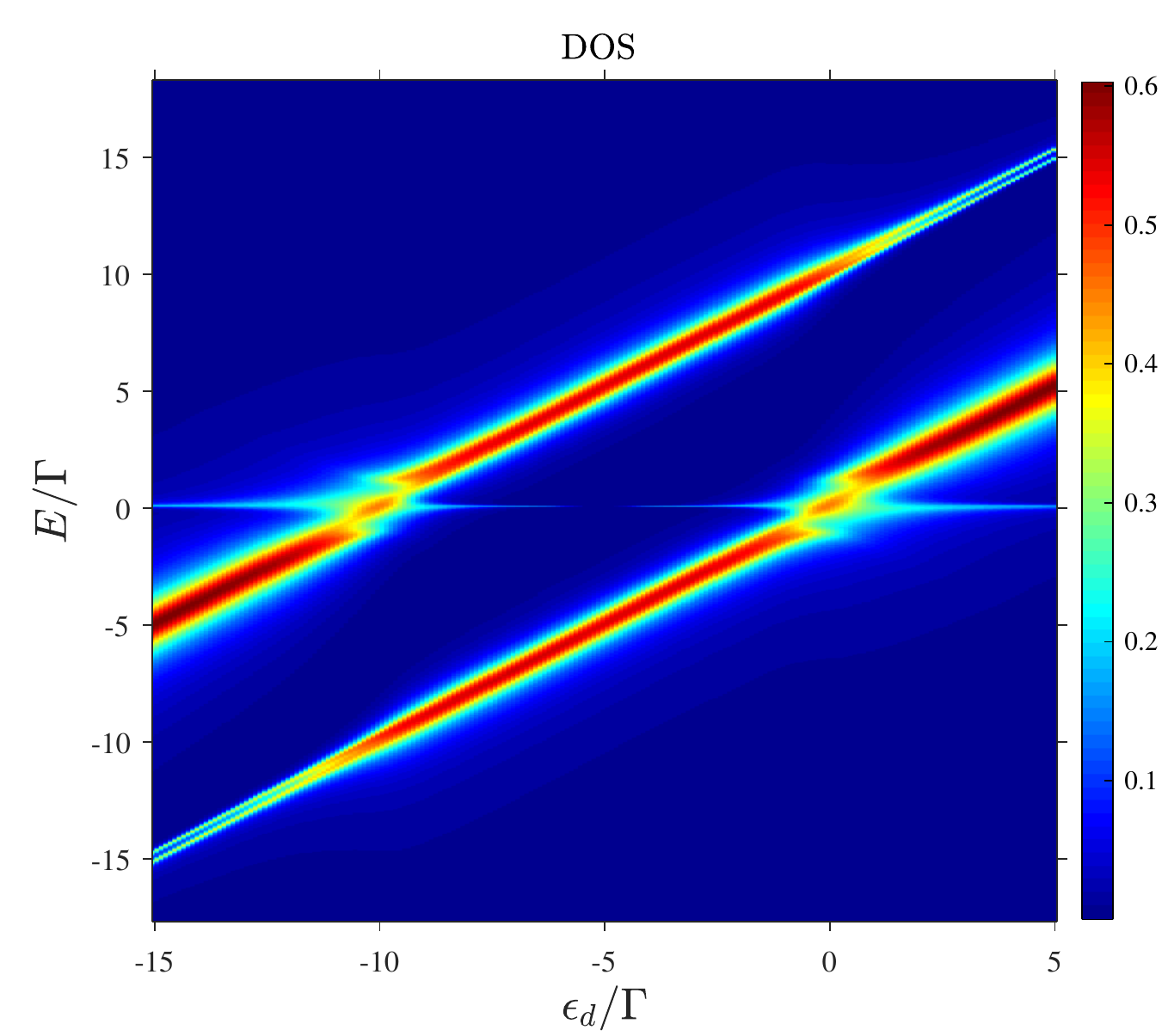}}
		\subfigure[][]{\includegraphics[width=3.4cm]{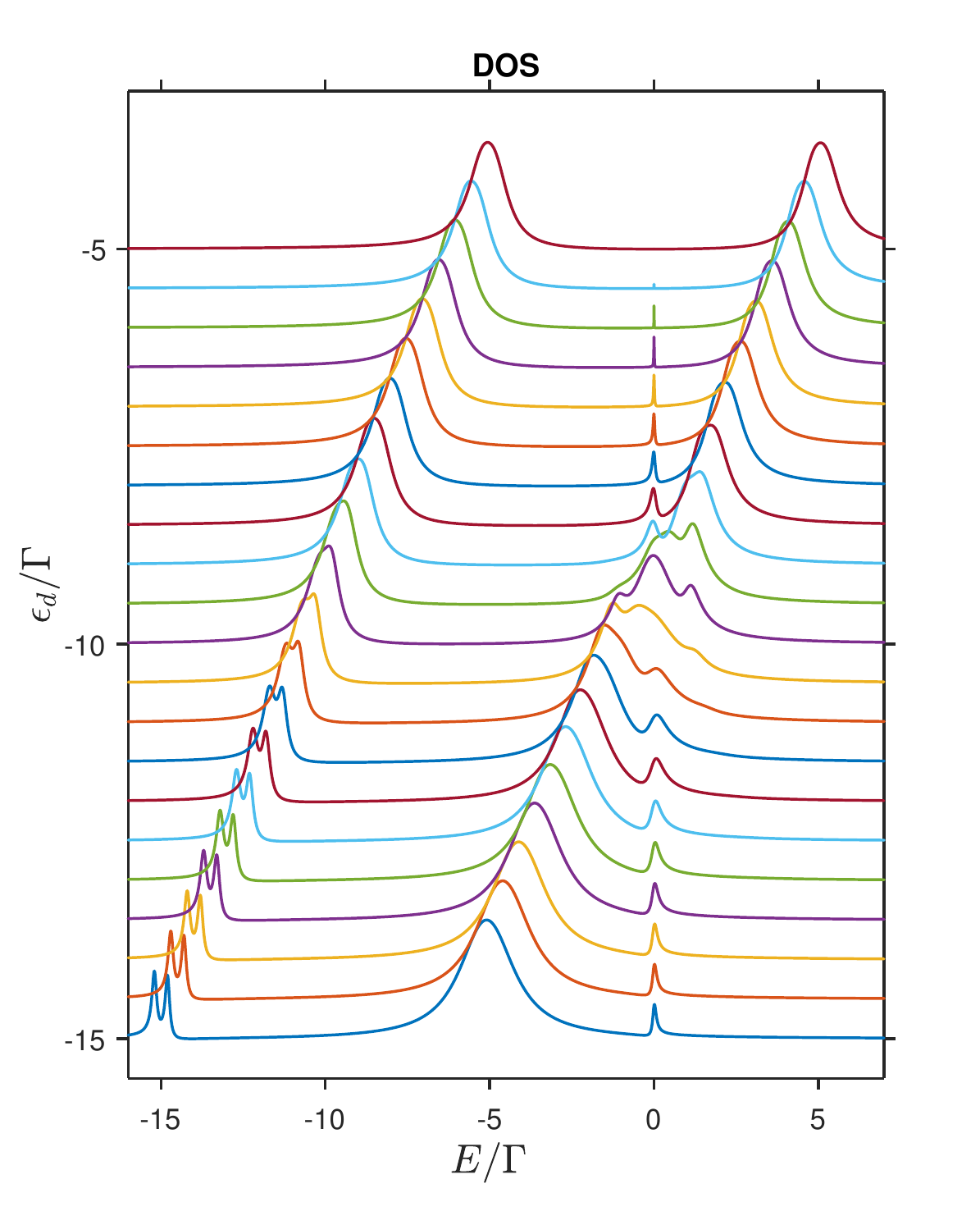}}
	\end{center}	
	\caption{In the case of zero MBS coupling ($\epsilon_M=0$), the DOS in QD as a function of (gate-controlled) QD energy level $\epsilon_d$. Parameters are $U=10\Gamma$, $\lambda=0.5\Gamma$, $E_z=0.4\Gamma$, $k_B T=0.1\Gamma$.}
	\label{fig:DOS_zero_epsilon}
\end{figure}

\begin{figure}
	\begin{center}
		\subfigure{\includegraphics[height=8.7cm]{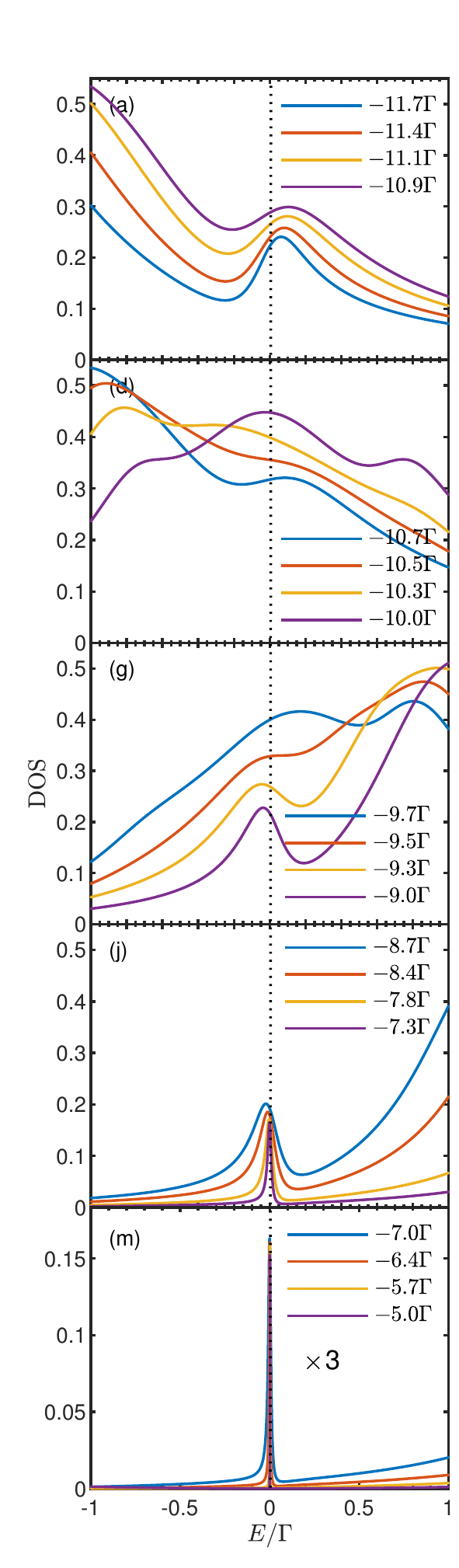}}
		\subfigure{\includegraphics[height=8.7cm]{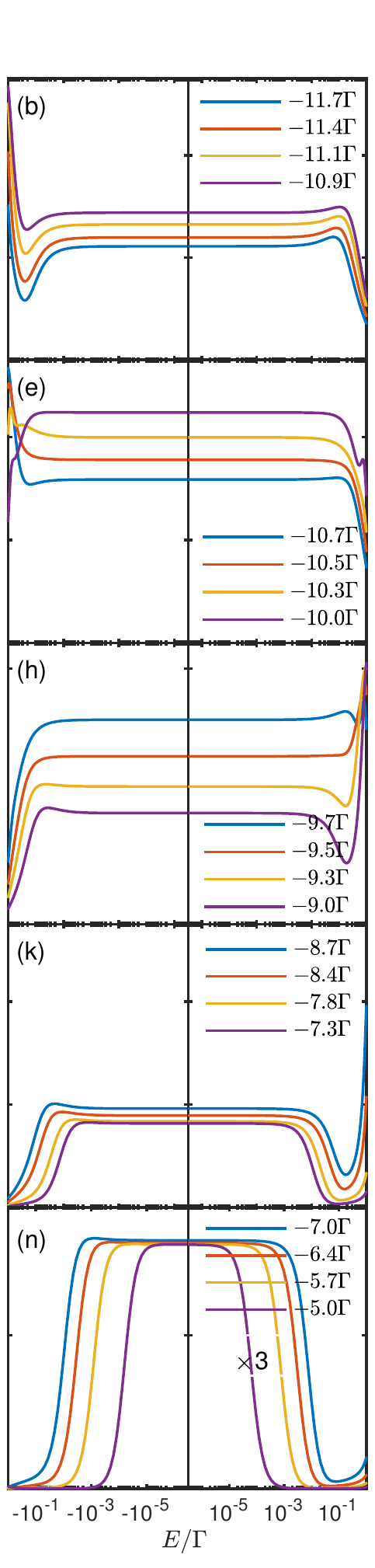}}
		\subfigure{\includegraphics[height=8.7cm]{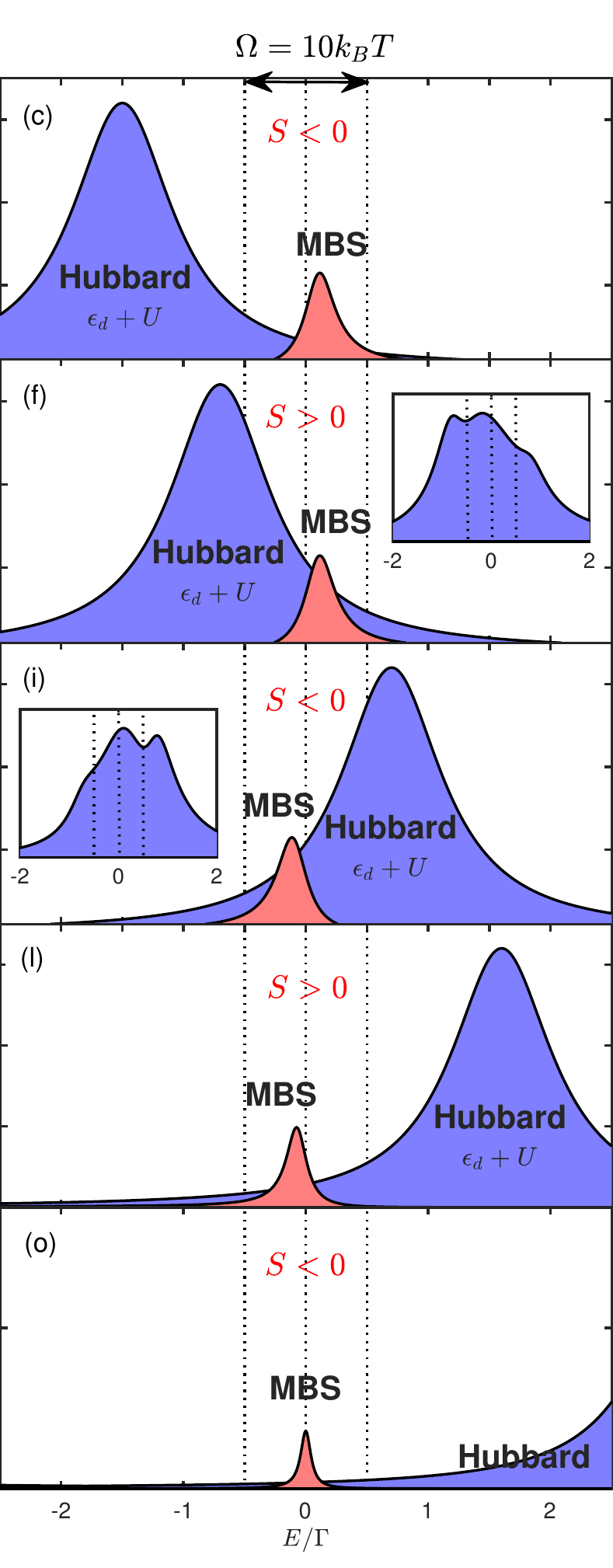}}
	\end{center}
	\caption{
		For $\epsilon_M=0$, the DOS near $E=0$. First and second columns plot the DOS for annotated values of $\epsilon_d$ by using a linear and logarithmic scale for the $E$-axis, respectively. Third column is the schematic diagram explaining the sign change of thermopower: The big and small peaks represent Hubbard and MBS peak, respectively, and dotted lines are the boundaries of the thermal activation window defined by the width $\Omega=10{k_B}T$. Inset in (f) and (i) show that the Hubbard and MBS peak are mixed and formed asymmetrical three peaks. The other parameters are the same as those in Fig. \ref{fig:DOS_zero_epsilon}.}
	\label{fig:DOS_zero_why}
\end{figure}

Fig. \ref{fig:DOS_zero_epsilon} shows the DOS in QD as a function of (gate-controlled) QD energy level $\epsilon_d$ in the case of zero MBS coupling ($\epsilon_M=0$).
As shown in Fig. \ref{fig:DOS_zero_epsilon}(a), DOS is symmetric about the particle-hole symmetric point ($2\epsilon_d+U=0$), $\epsilon_d=-5\Gamma$, and there are three peaks in DOS.
Two peaks (Hubbard peaks) appear near the effective energy levels in QD ($E=\epsilon_d$ and $E=\epsilon_d+U$), while smaller one of them is split by $E_z=0.4\Gamma$ due to the Zeeman splitting and larger one isn't split.
Such a splitting becomes weaker and weaker and finally disappears when their weights are nearly same (The reason is why $E_z=0.4\Gamma$ is smaller than the QD-lead coupling $\Gamma$).
On other hand, very small peak (MBS peak) appears at $E=0$, which is concerned about existence of MBS [see Fig. \ref{fig:DOS_zero_epsilon}(b)]. 
When the energy level in QD is approached to the chemical potential of the leads $\mu=0$ ($\epsilon_d=-10\Gamma$), the Hubbard peak and MBS peak are mixed and formed three peaks (for small $\lambda$, these may be formed two-peak structure \cite{Liu_PRB84}), and at exactly $\epsilon_d=-10\Gamma$, these peaks become symmetrical.
Furthermore, it is important that MBS peak near $E=0$ leans to the right (left) if the neighbour Hubbard peak is on the left (right), and becomes weak in the vicinity of the $\epsilon_d=-5\Gamma$ (see Fig. \ref{fig:DOS_zero_why}).

\begin{figure}
	\begin{center}
		\subfigure[ ]{\includegraphics[width=4.7cm]{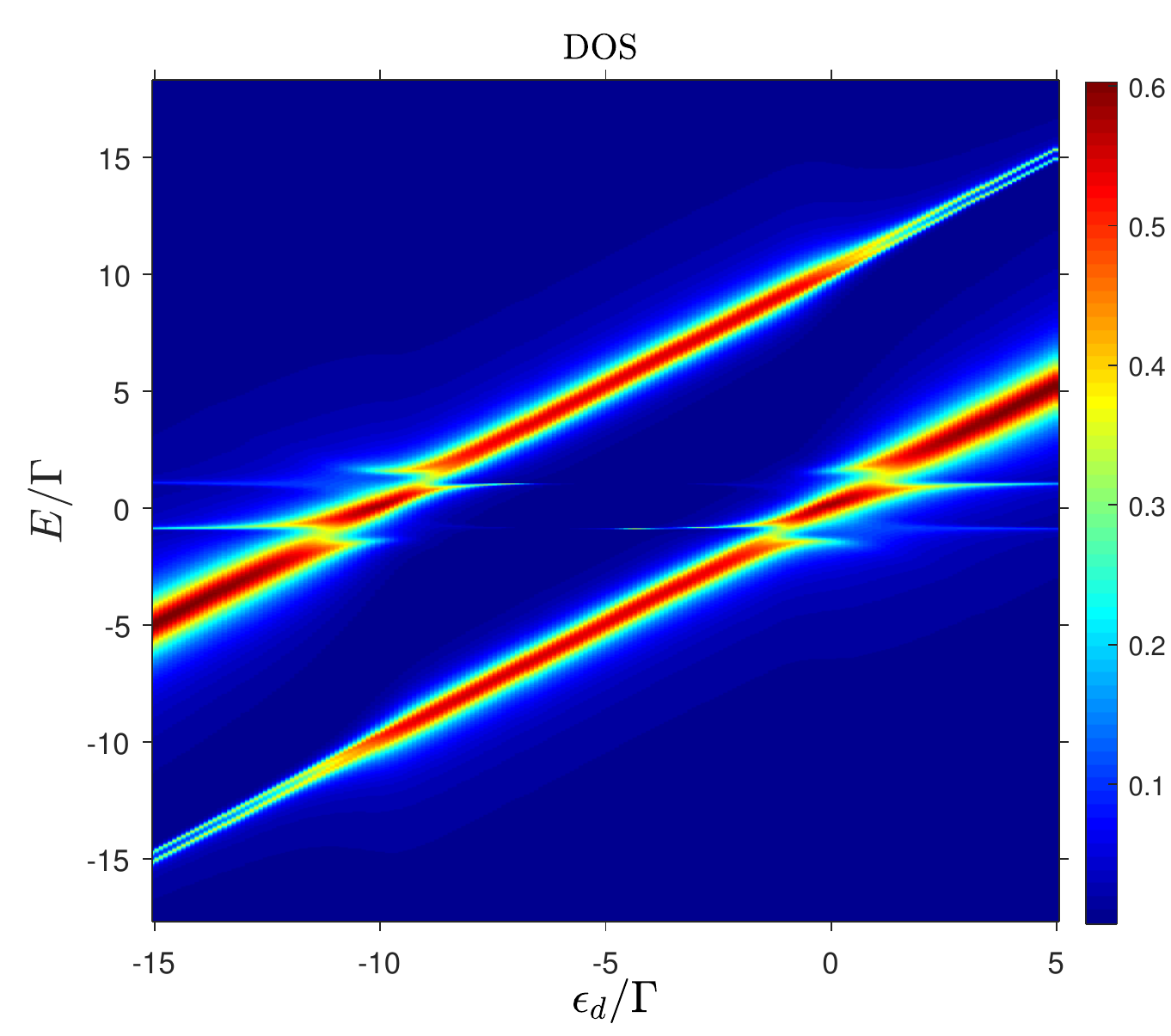}}
		\subfigure[ ]{\includegraphics[width=3.4cm]{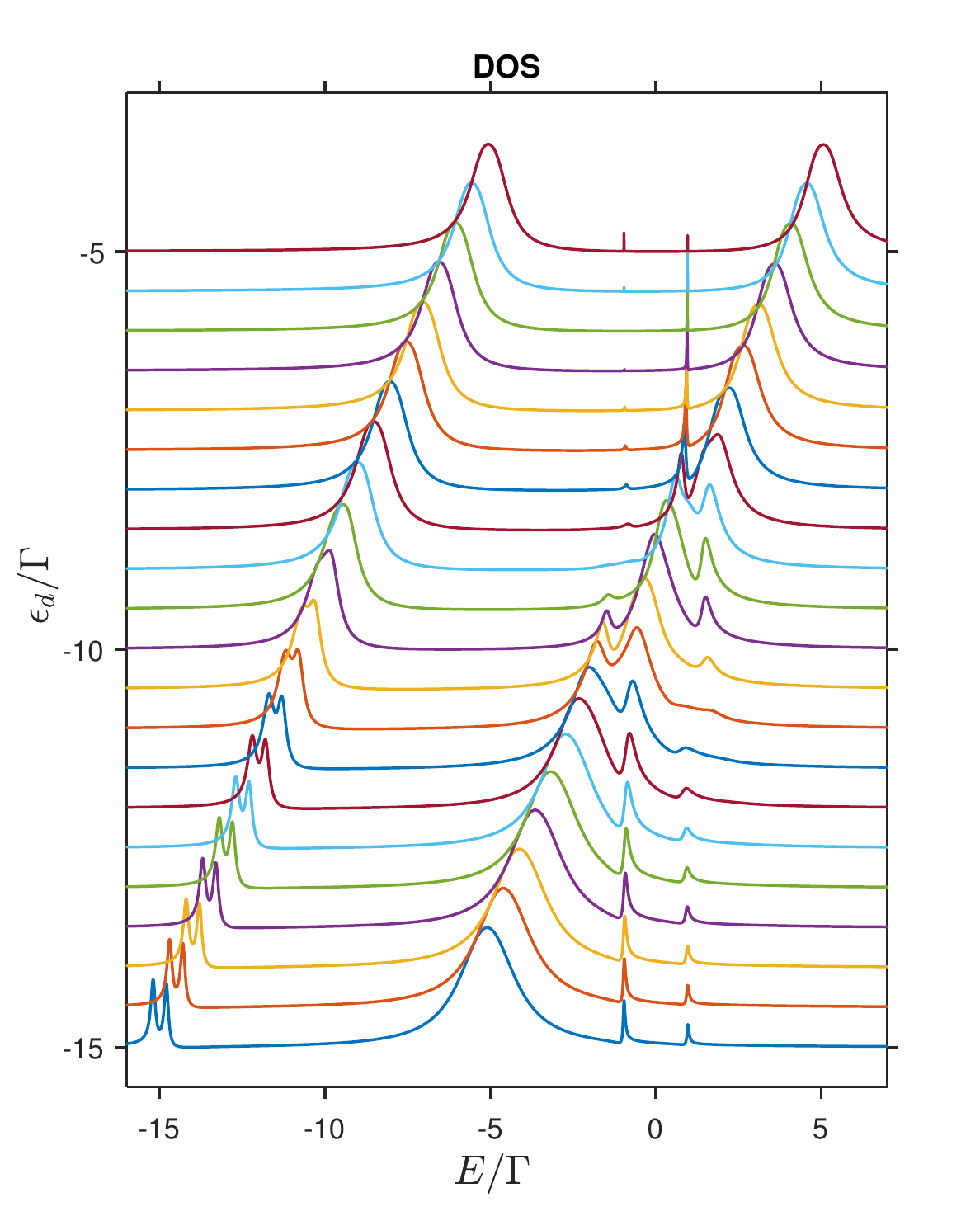}}
	\end{center}
	\caption{
		In the case of nonzero MBS coupling ($\epsilon_M=0.5\Gamma$), the DOS in QD as a function of (gate-controlled) QD energy level $\epsilon_d$. The other parameters are the same as those in Fig. \ref{fig:DOS_zero_epsilon}.
	}
	\label{fig:DOS_nonzero_epsilon}
\end{figure}

However, for $\epsilon_M\neq 0$, the characteristics of MBS peak in DOS shows a striking difference for $\epsilon_M=0$ mentioned above.
In case of $\epsilon_M = 0.5\Gamma$, the DOS in QD as a function of $\epsilon_d$ has been shown in Fig. \ref{fig:DOS_nonzero_epsilon}.
The positions and heights of the Hubbard peaks are nearly the same with one's for $\epsilon_M=0$ [see Fig. \ref{fig:DOS_nonzero_epsilon}(a)].
But two MBS peaks appear at $E=\pm2\epsilon_M$ and their heights are asymmetrical due to the neighbour Hubbard peaks, while they are symmetric at the position $\epsilon_d=-5\Gamma$ [see Fig. \ref{fig:DOS_nonzero_epsilon}(b)].
Just as in the case of $\epsilon_M=0$, when the energy level in QD approaches to the chemical potential of the leads $\mu=0$, the Hubbard peak and MBS peak are mixed and formed asymmetrical three peaks (for small $\lambda$, two peaks), and exactly at the position $\epsilon_d=-10\Gamma$, these peaks become symmetrical.

%\subsection{Characteristics of Thermoelectricity}

\begin{figure}
\begin{center}
  \includegraphics[width=8.0cm]{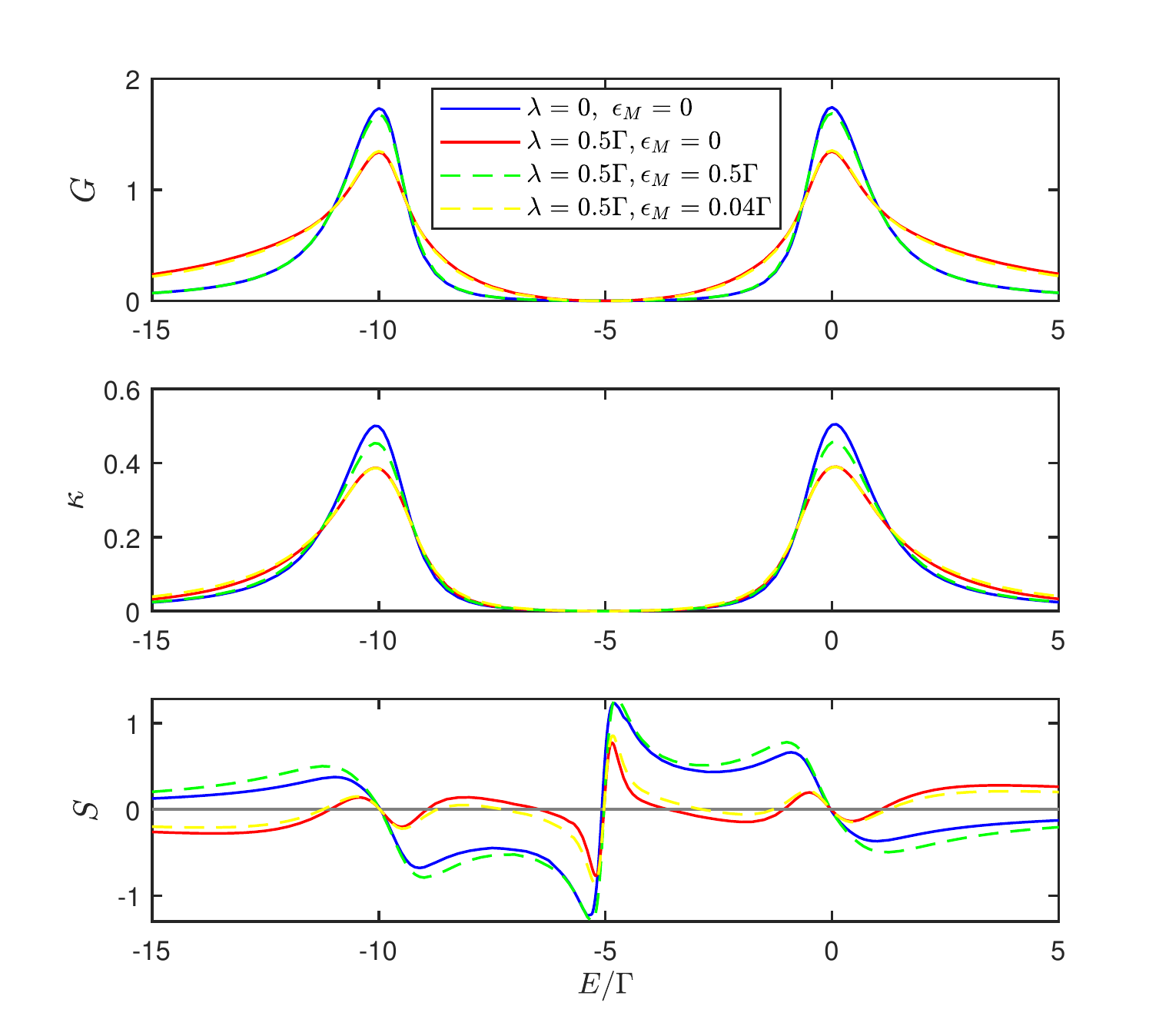}
\end{center}
	\caption{
		The electric conductance $G(e^2/h)$, the thermal conductance $\kappa(k_B/h)$ and the thermopower $S(k_B/e)$ as a function of $\epsilon_d$ for different $\epsilon_M$. The other parameters are the same as those in Fig. \ref{fig:DOS_zero_epsilon}.
	}
	\label{fig:thermo_charact}
\end{figure}

%
Such complicated properties of DOS affect the thermoelectric characteristics.
The characteristics of thermoelectric transport shows very special modality due to the presence of MBS and Coulomb interaction in QD.
Fig. \ref{fig:thermo_charact} shows the electric conductance $G$, the thermal conductance $\kappa$ and the thermopower $S$ as a function of $\epsilon_d$ for different $\epsilon_M$. 

The electric conductance $G$ is symmetric about $\epsilon_d=-5\Gamma$ due to the particle-hole symmetry and there are two resonant peaks when the two effective energy levels in QD fit with Fermi level of leads.
In case that QD is coupled to ideal isolated MBS ($\lambda=0.5\Gamma$, $\epsilon_M=0$), the height of resonant peak reduces by about 3/4 than the one without MBS ($\lambda\neq0$), which is coincided with the result in previous study \cite{Lee_PRB87}.
For $\epsilon_M=0.5\Gamma$, the properties of $G$ is nearly same with the case for one without MBS. 
The behaviour of thermal conductance $\kappa$ is similar to $G$ except for quantitative differences.

The thermopower $S$ shows very fantastic manner.
At first, for $\lambda=0$, the sign of $S$ changes at three points: one is the particle-hole symmetric point, while the others are the points where either of energy levels in QD is fitted with Fermi level of leads.
For $\lambda=0.5\Gamma$, $\epsilon_M=0$, the sign of $S$  changes 9 times, including above three times (Note that at above three points, $S$ has the same tangent for $\lambda=0$).
The reason why sign of $S$ behaves complicatedly is that the MBS peak near $E=0$ leans to the left or right according to the changes of $\epsilon_d$, due to the shifting effects by interacting with two QD levels (see Fig. \ref{fig:DOS_zero_why}).
The sign of thermopower $S$ is associated with behaviour of DOS near $E=0$ \cite{Dong_JPCM14} [see Eqs. (\ref{eq:def_I}), (\ref{eq:charact_th})]. 
At low temperature the derivation of Fermi-Dirac distribution function by the energy ${f}'(E)$ forms a negative sharpen peak, the full width at half maximum of the peak is $3.5k_B T$, hence ${f}'(E)$ defines a thermal activation window with the width $\Omega =10{k_B}T$.
Note that the DOS outside this window has almost no contribution to $S$ \cite{Ye_PRB90}.
As shown in third column of Fig. \ref{fig:DOS_zero_why}, MBS peak near $E=0$ locates within the thermal activation window and leans to the right (left) when the neighbour Hubbard peak locates on the left (right).
In Fig. \ref{fig:DOS_zero_why}(c), the Hubbard peak at $E=\epsilon_d+U$ stays outside the window (its tail may extend into the window), whereas MBS peak near $E=0$ leans to the right, therefore, $S<0$ (electron-like).
However, when the MBS peak and the tail of Hubbard peak within the window contribute equally to thermopower, $S=0$.  
If the Hubbard peak lies partly within the window [Fig. \ref{fig:DOS_zero_why}(f)], its contribution to $S$ is larger than that of the MBS peak, leading to a positive $S$ (hole-like).
When $\epsilon_d+U$ is approached to $\mu=0$, the Hubbard and MBS peaks are mixed and formed asymmetric three peaks (for small $\lambda$, two peaks) going down to the right [see inset in Fig. \ref{fig:DOS_zero_why}(f)], therefore $S>0$.
Note that exactly at the position $\epsilon_d=-10\Gamma$, these peaks become symmetrical, leading to a zero $S$.
The situations in Fig. \ref{fig:DOS_zero_why}(i) and (l) are reversed from that in Fig. \ref{fig:DOS_zero_why}(f) and (c), leading to negative and positive $S$, respectively.
Note that in the former case, the MBS leans to the left weaker than in the latter case and its width becomes smaller, because the MBS peak lies between the two Hubbard peaks at $E=\epsilon_d$, $E=\epsilon_d+U$.
As $\epsilon_d$ approaches the particle-hole symmetry point, the width of the MBS peak becomes smaller and the MBS peak comes to be more symmetrical [Fig. \ref{fig:DOS_zero_why}(o)].
Therefore, the MBS peak has almost no contribution to $S$, whereas the small tail of the Hubbard peak located at the right gives the negative contribution, leading to a negative $S$.
When $\epsilon_M=0.5\Gamma$, however, the sign change in $S$ is nearly same for $\lambda=0$, because two MBS peaks appear not near $E=0$, but at $E=\pm 2\epsilon_M$ [see Fig. \ref{fig:DOS_nonzero_epsilon}(b)].
Among the two MBS peaks, the MBS peak placed on the side of neighbour Hubbard peak is larger than the other, therefore, they do not contribute to the $S$-sign.
To emphasize that even in case of $\epsilon_M=0.04\Gamma$, the sign of $S$ also changes 9 times due to the overlap of two MBS peaks at $E=0$.
That overlap becomes smaller and smaller according to the increasing of $\epsilon_M$, hence, original properties for $\lambda=0$ will be recovered.

\begin{figure*}[t!]
	\begin{center}
		\subfigure[$k_B T=0.05\Gamma$, $U=8\Gamma$]{\includegraphics[width=3.5cm]{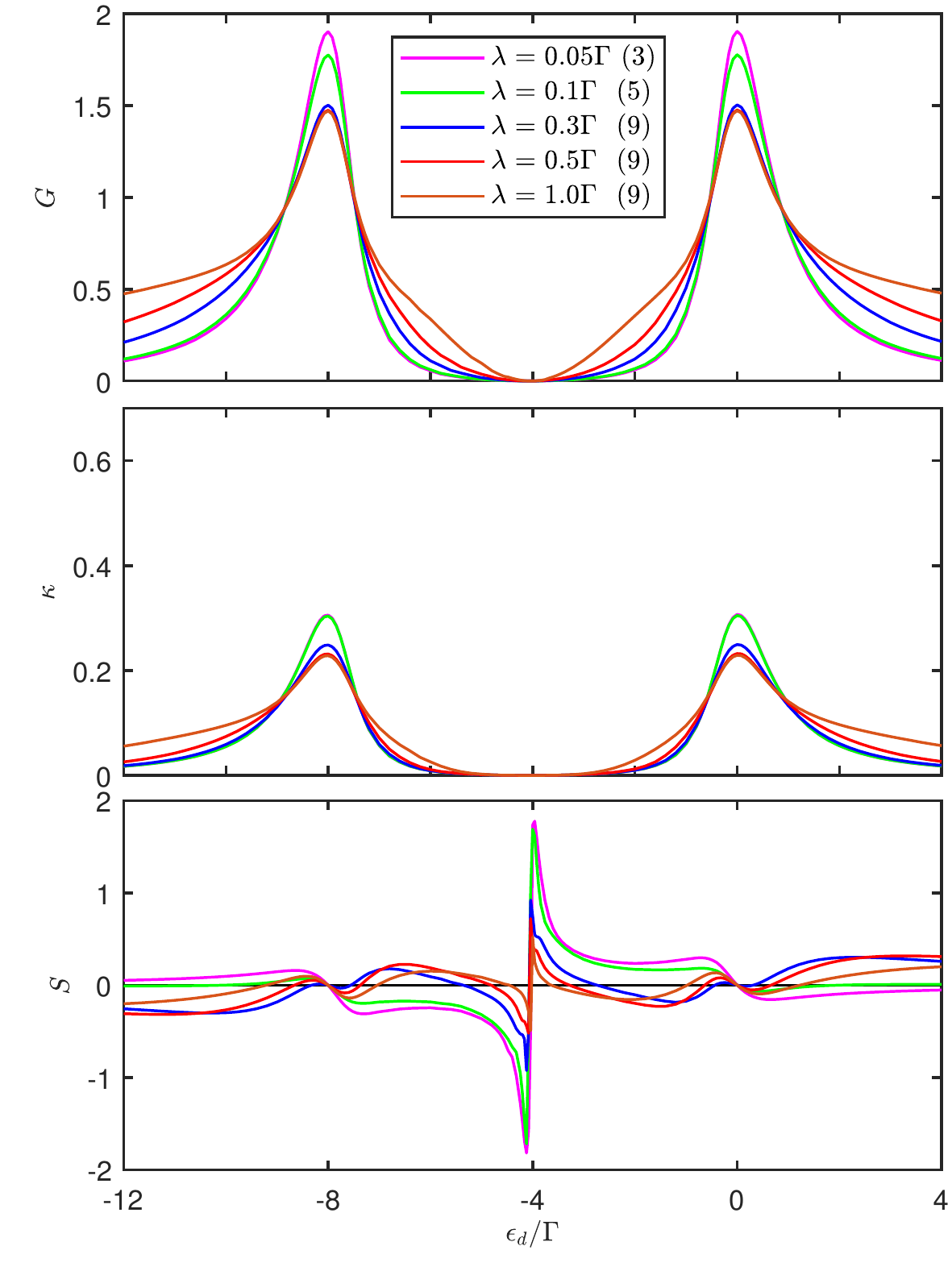}}
		\subfigure[$k_B T=0.05\Gamma$, $U=10\Gamma$]{\includegraphics[width=3.5cm]{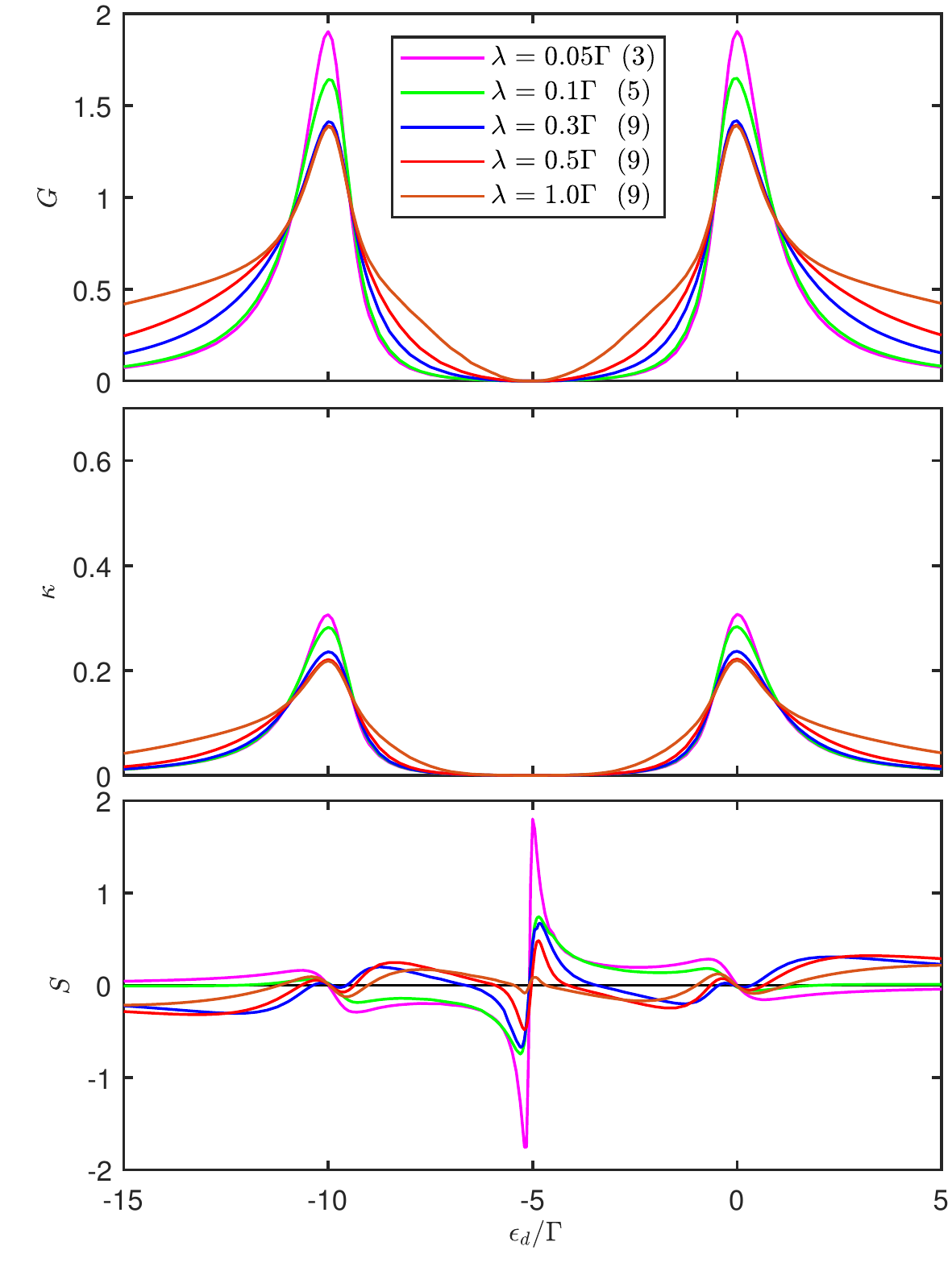}}
		\subfigure[$k_B T=0.05\Gamma$, $U=15\Gamma$]{\includegraphics[width=3.5cm]{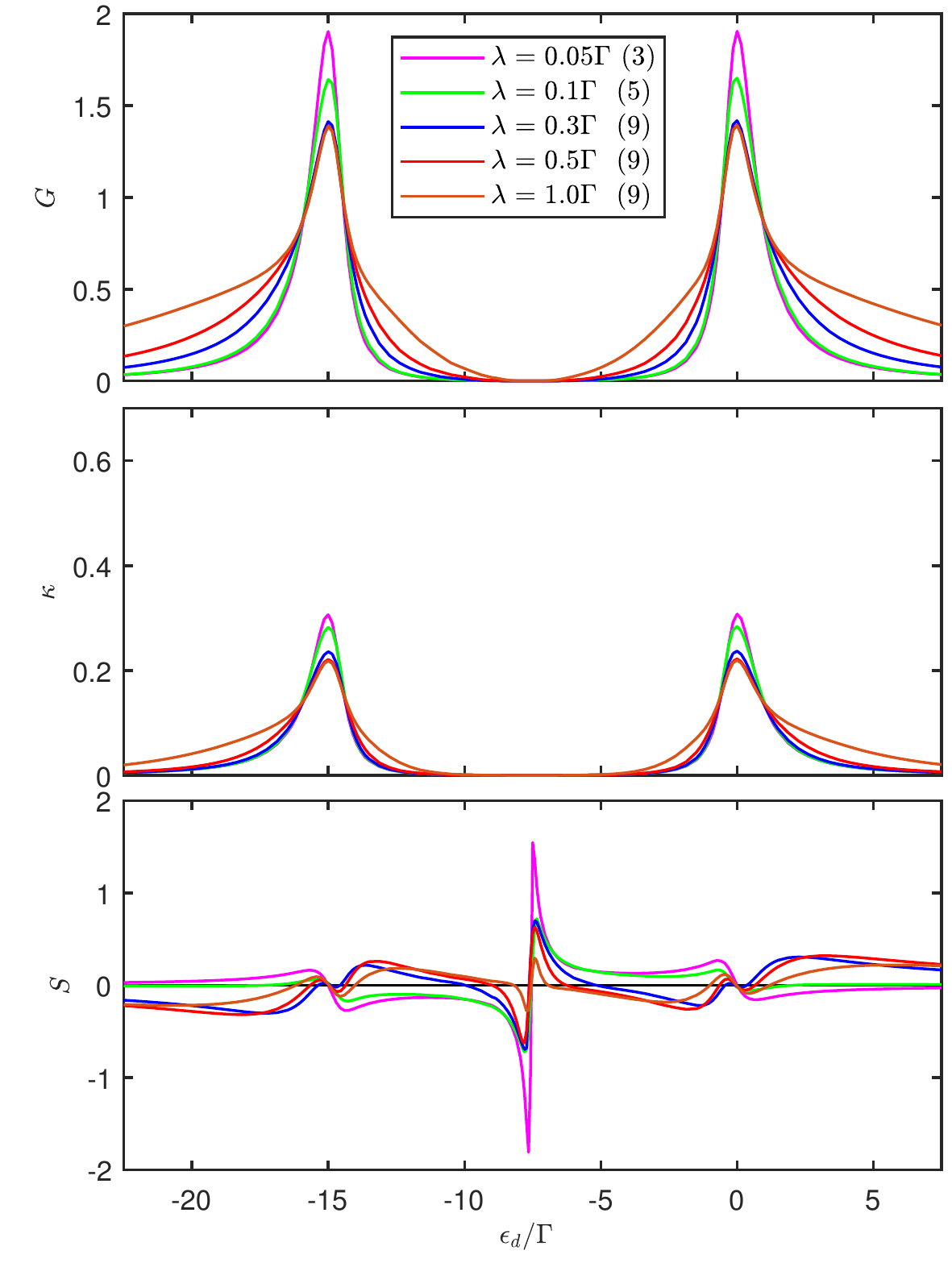}}
		\subfigure[$k_B T=0.05\Gamma$, $U=20\Gamma$]{\includegraphics[width=3.5cm]{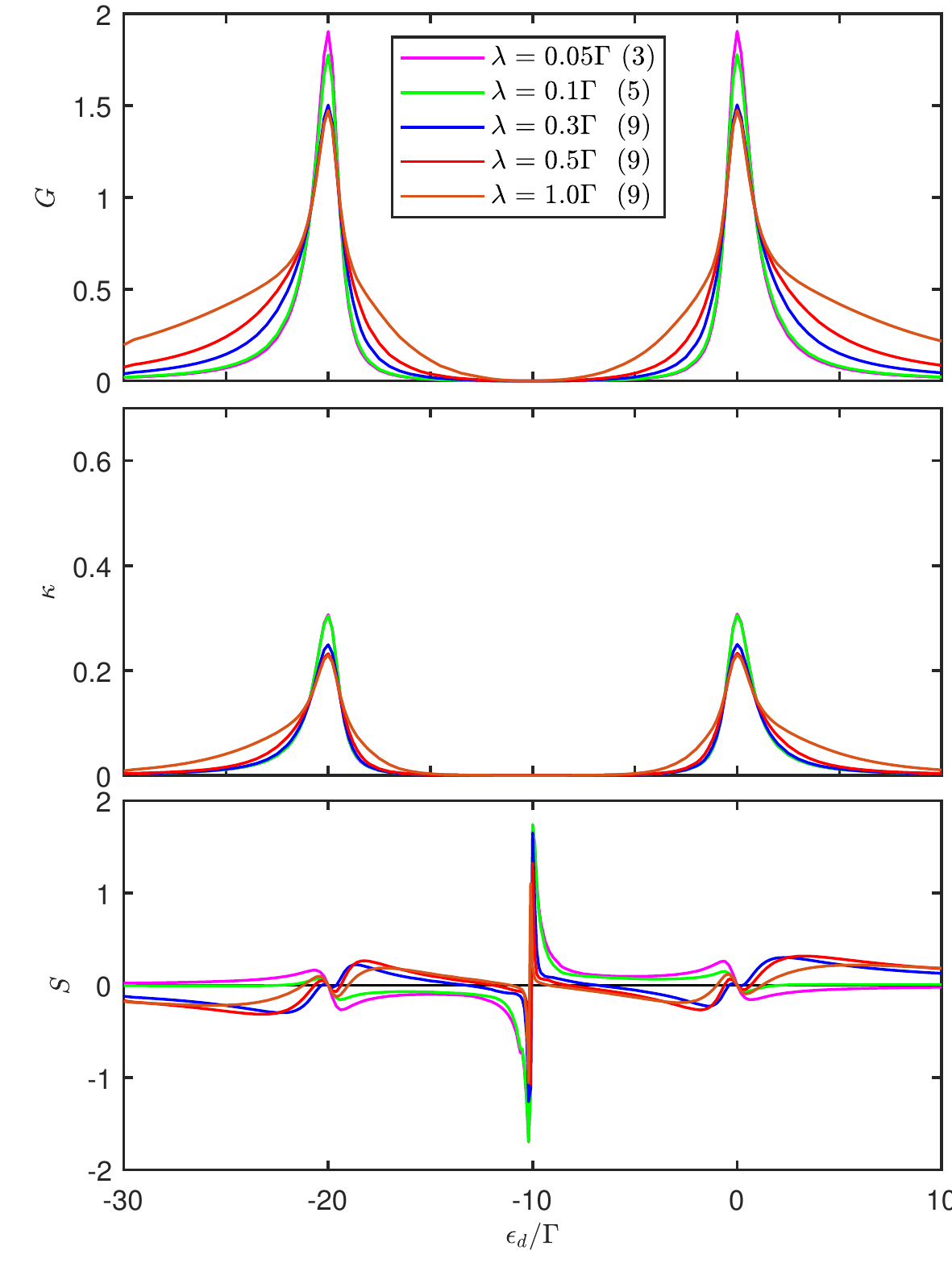}} \\
		\subfigure[$k_B T=0.07\Gamma$, $U=8\Gamma$]{\includegraphics[width=3.5cm]{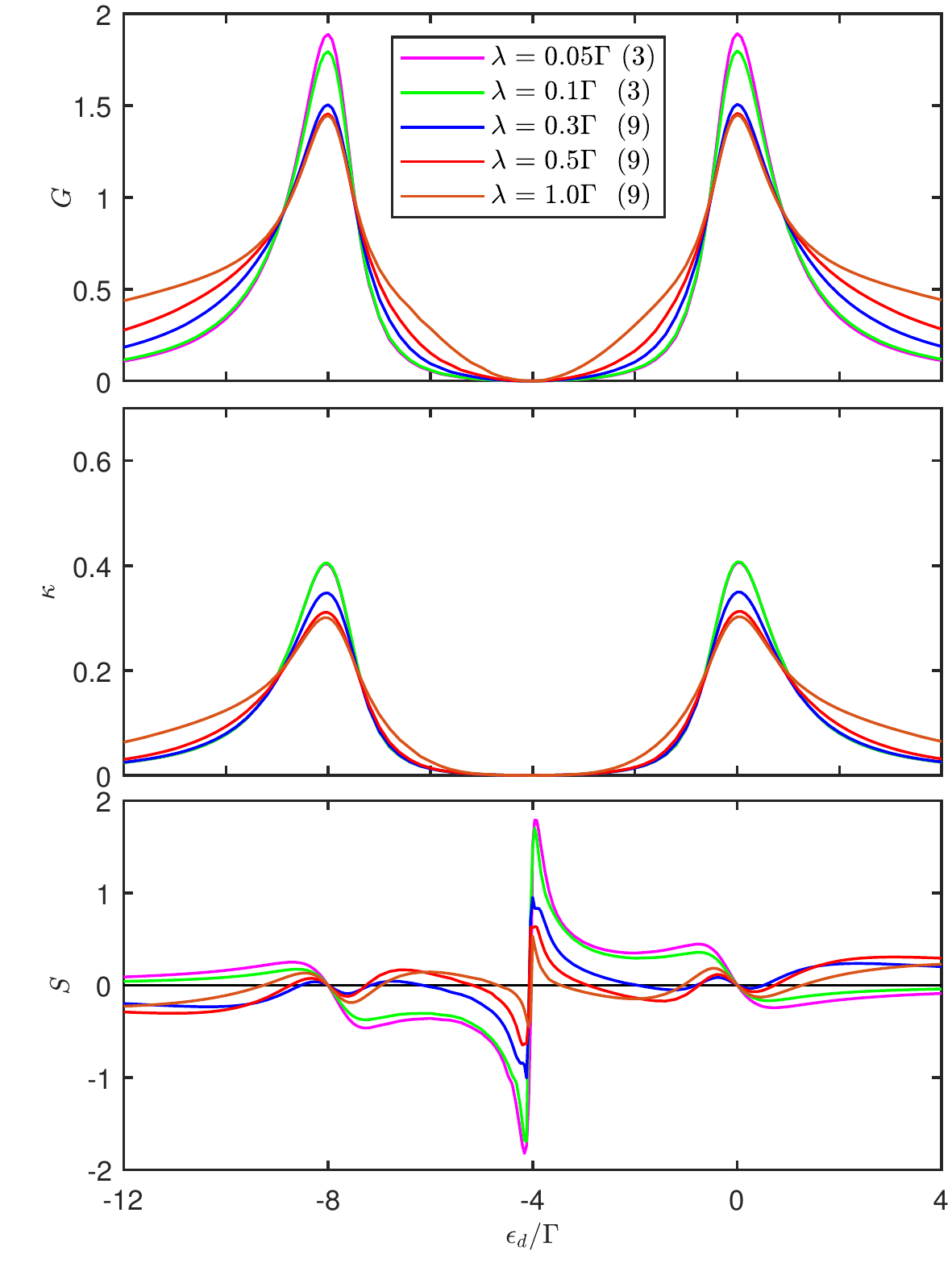}}
		\subfigure[$k_B T=0.07\Gamma$, $U=10\Gamma$]{\includegraphics[width=3.5cm]{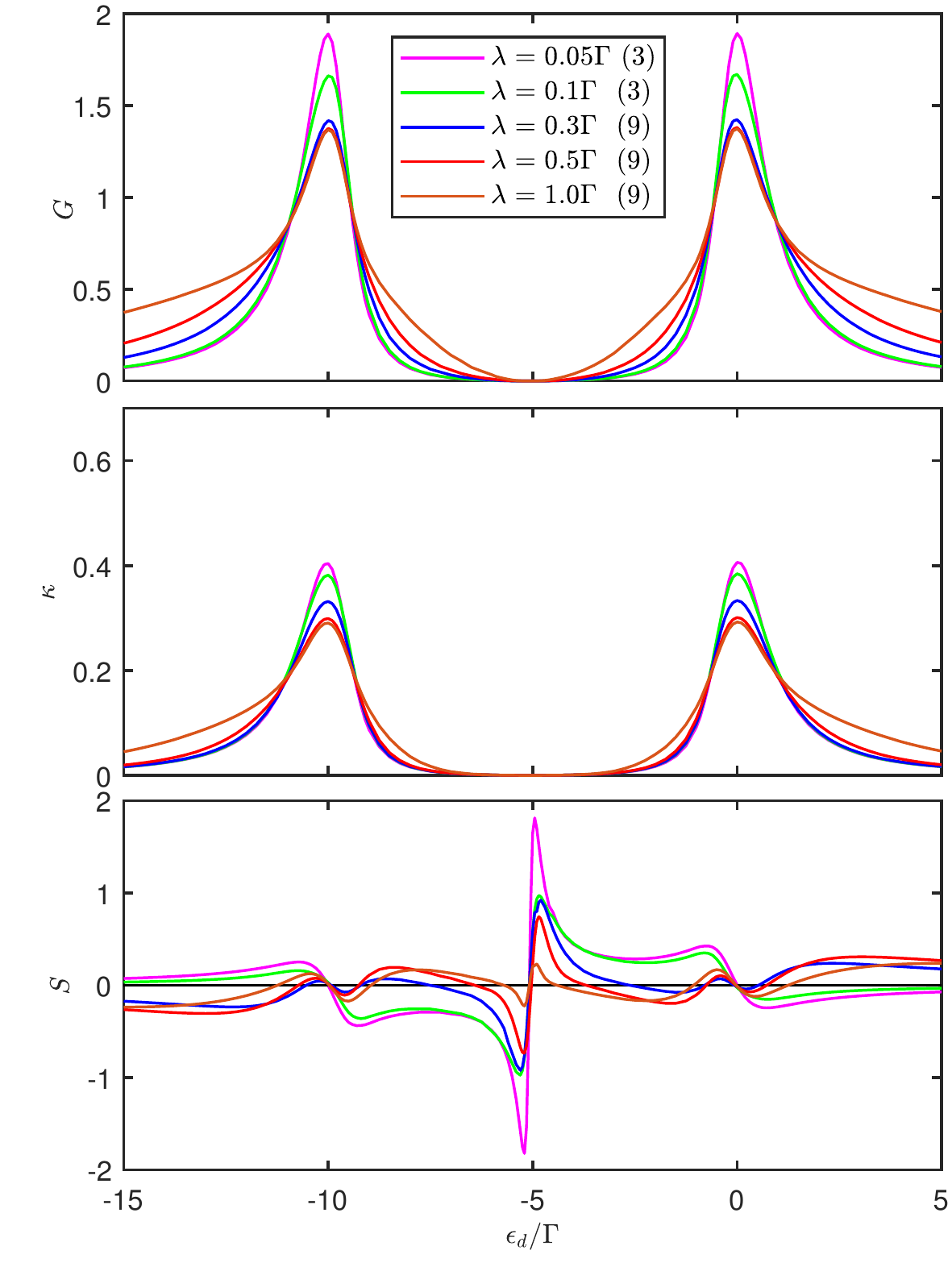}}
		\subfigure[$k_B T=0.07\Gamma$, $U=15\Gamma$]{\includegraphics[width=3.5cm]{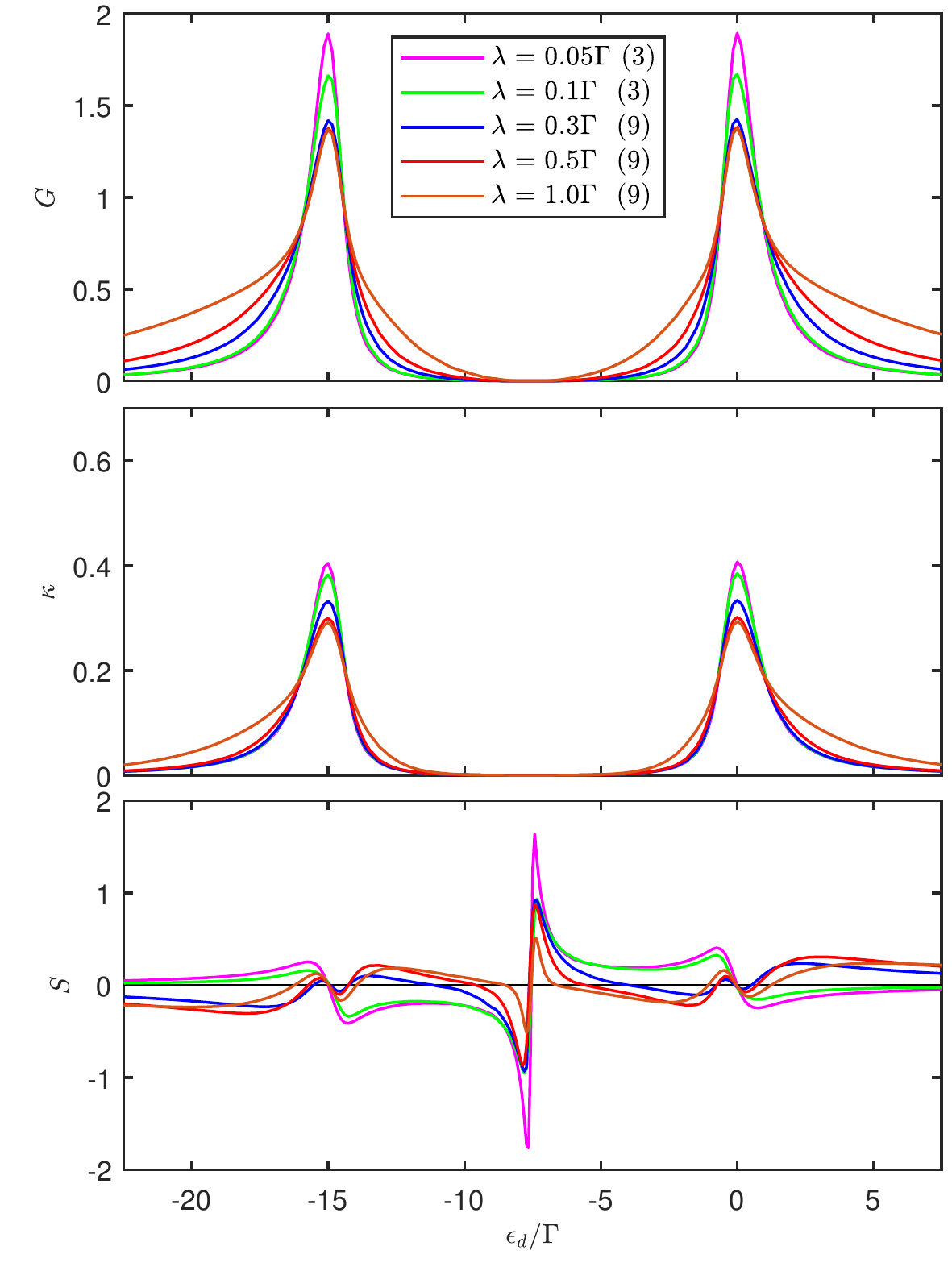}}
		\subfigure[$k_B T=0.07\Gamma$, $U=20\Gamma$]{\includegraphics[width=3.5cm]{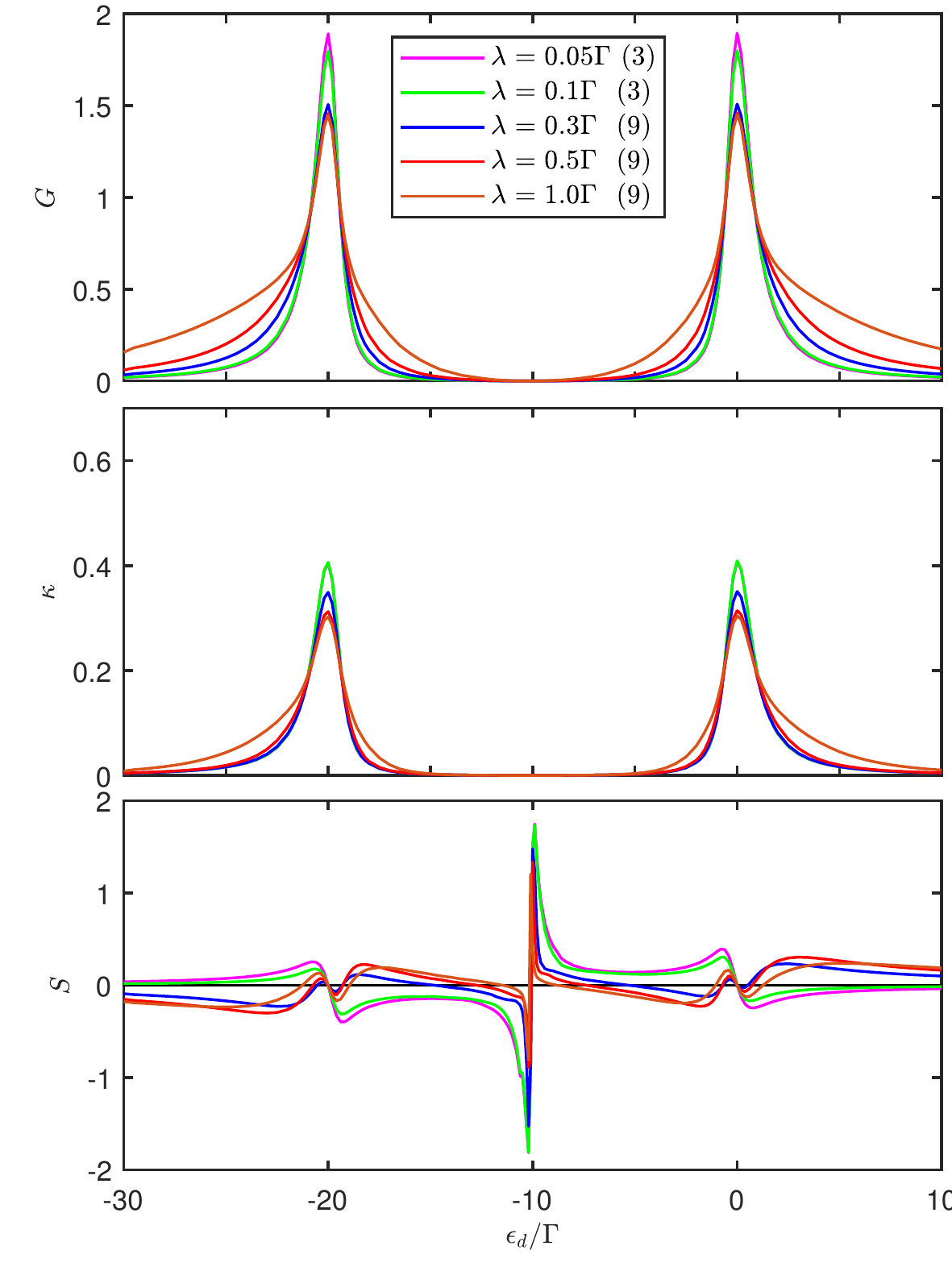}} \\
		\subfigure[$k_B T=0.1\Gamma$, $U=8\Gamma$]{\includegraphics[width=3.5cm]{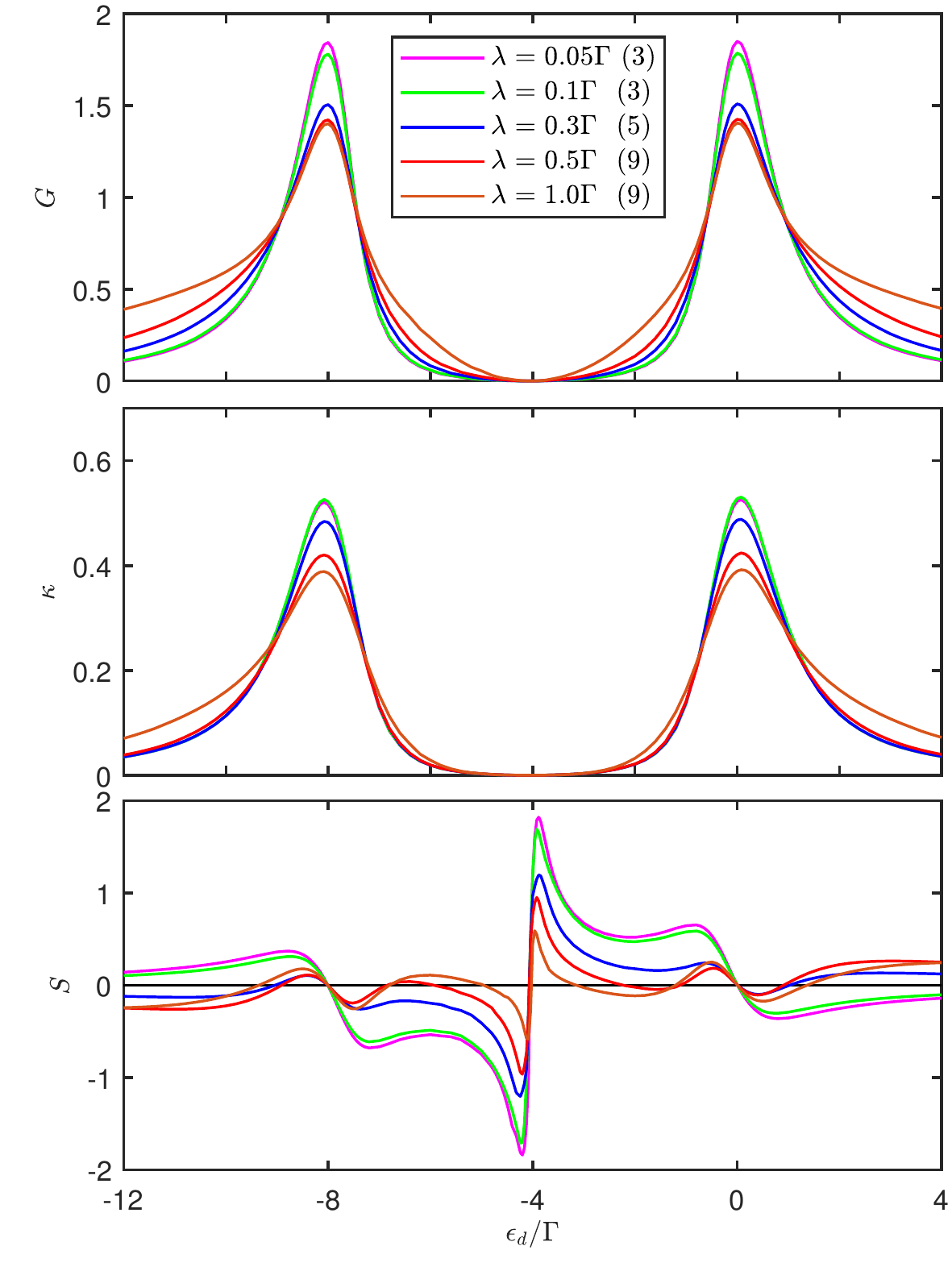}}
		\subfigure[$k_B T=0.1\Gamma$, $U=10\Gamma$]{\includegraphics[width=3.5cm]{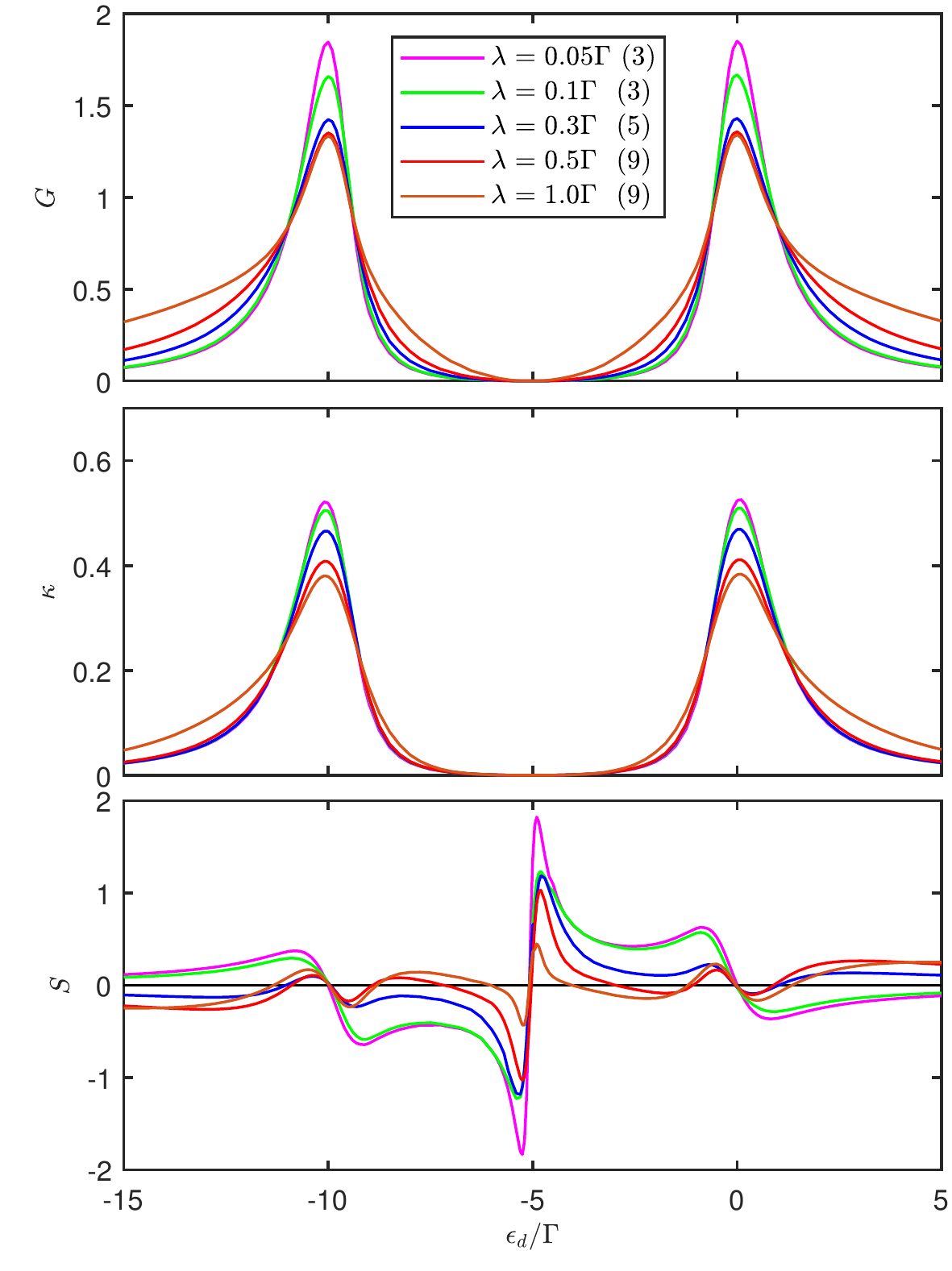}}
		\subfigure[$k_B T=0.1\Gamma$, $U=15\Gamma$]{\includegraphics[width=3.5cm]{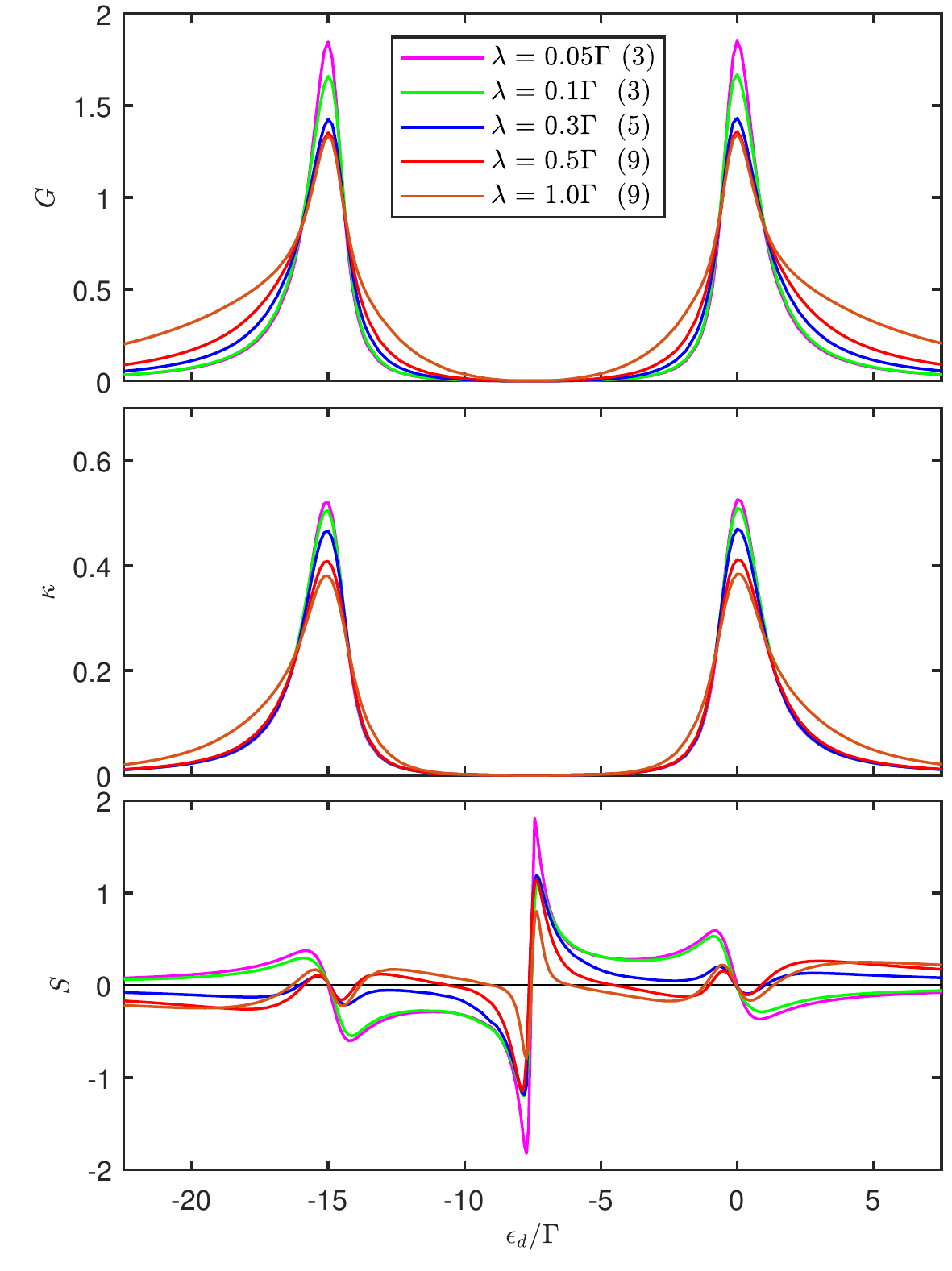}}
		\subfigure[$k_B T=0.1\Gamma$, $U=20\Gamma$]{\includegraphics[width=3.5cm]{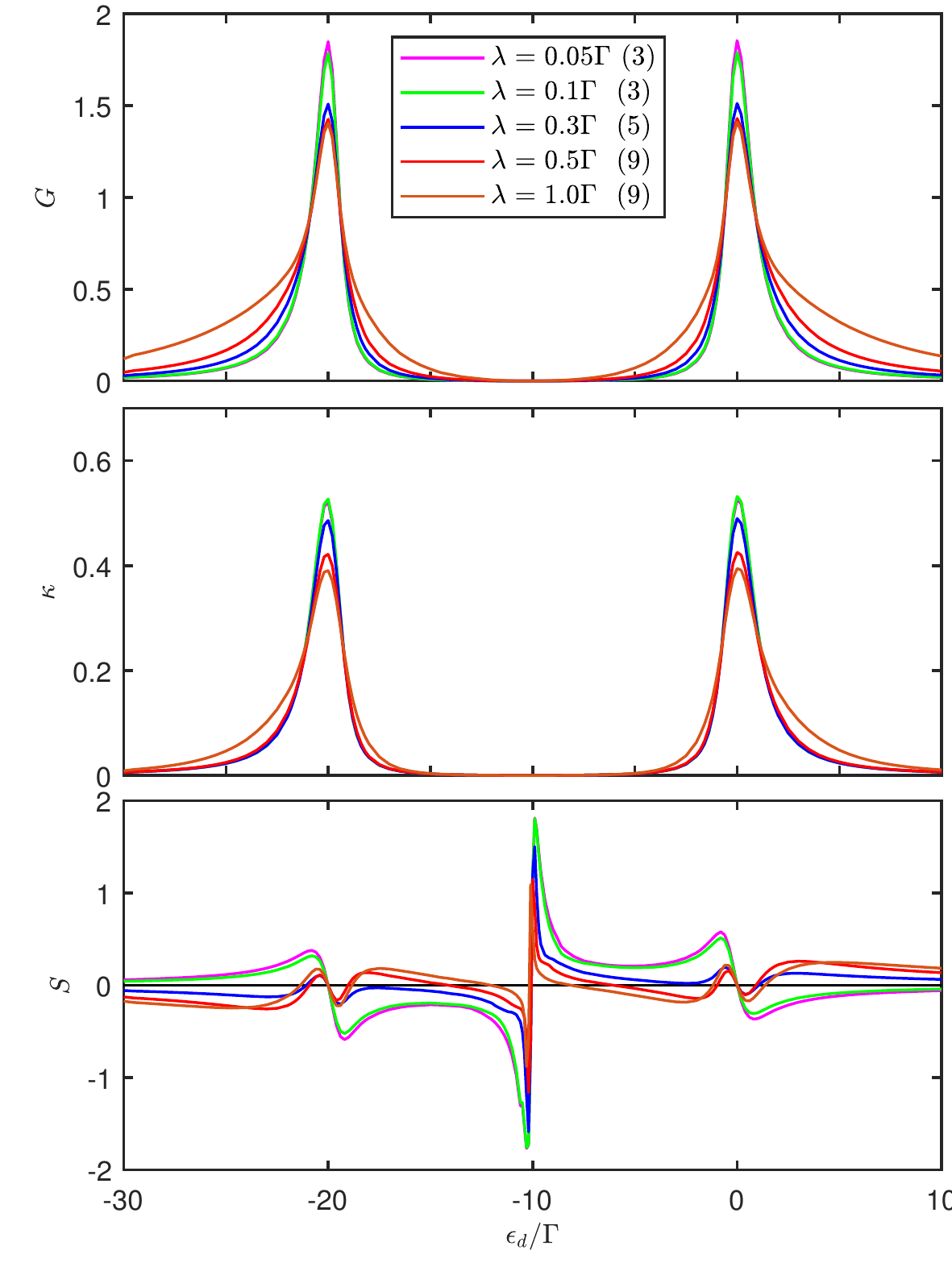}} \\
		\subfigure[$k_B T=0.15\Gamma$, $U=8\Gamma$]{\includegraphics[width=3.5cm]{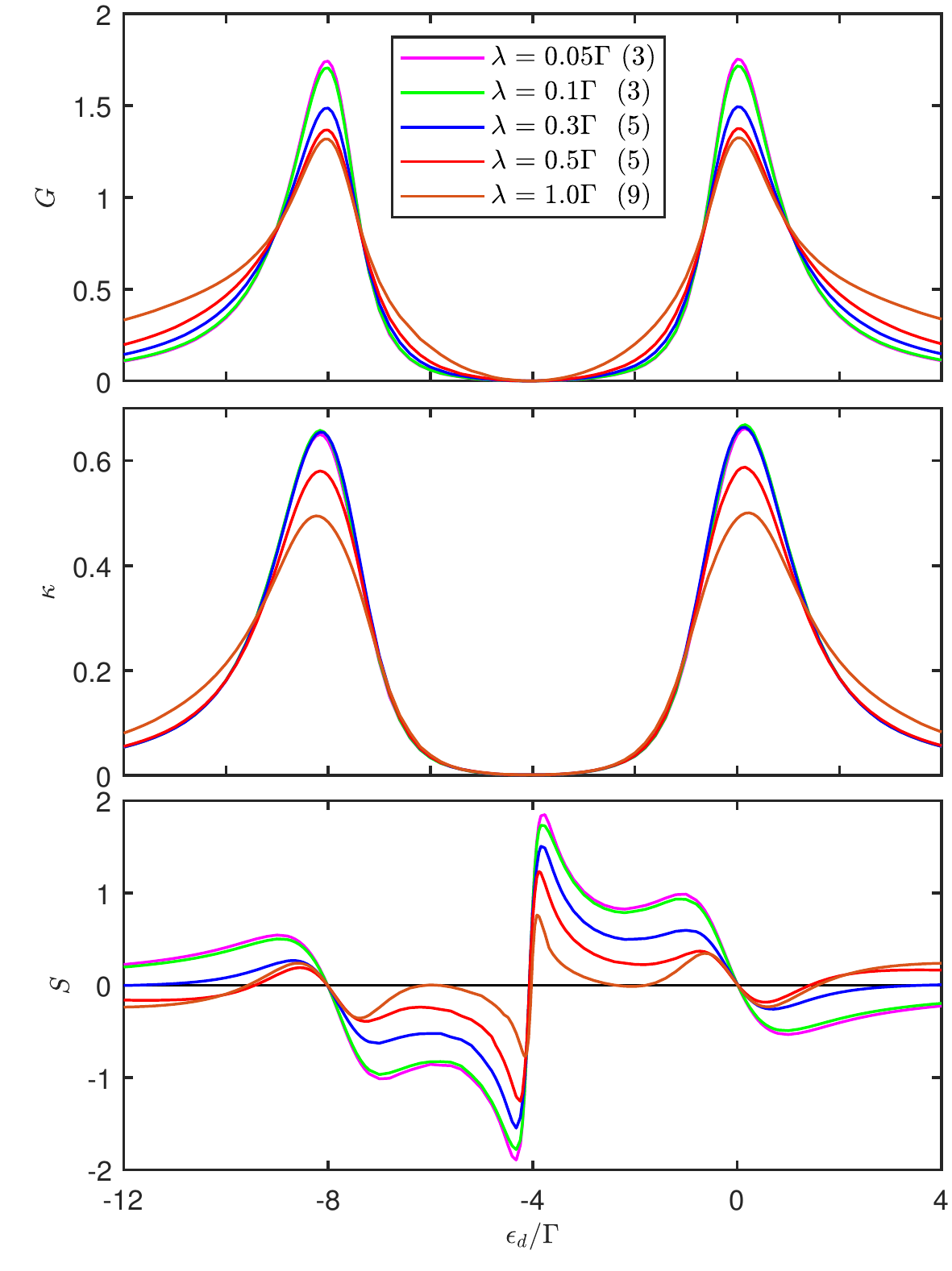}}
		\subfigure[$k_B T=0.15\Gamma$, $U=10\Gamma$]{\includegraphics[width=3.5cm]{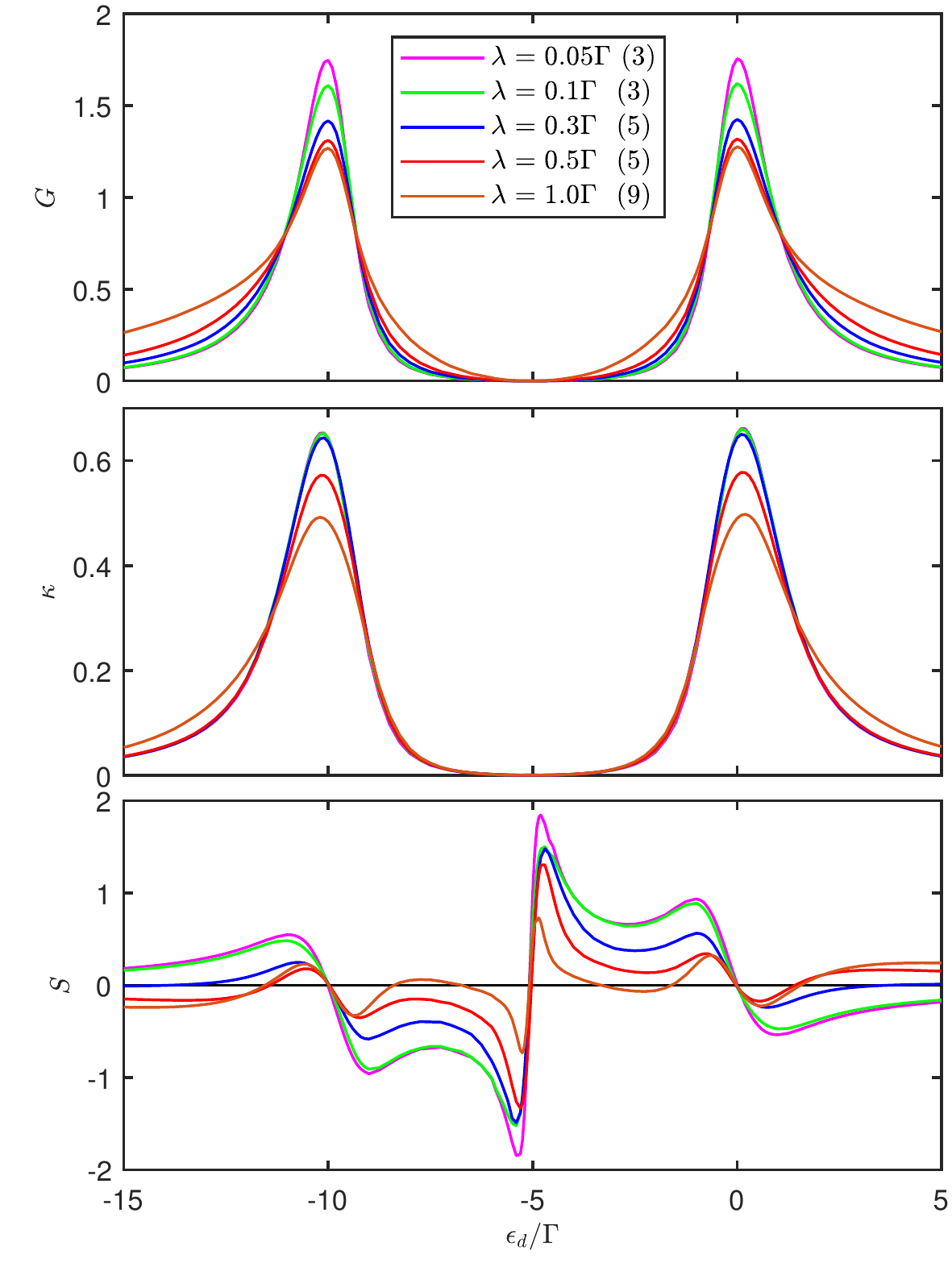}}
		\subfigure[$k_B T=0.15\Gamma$, $U=15\Gamma$]{\includegraphics[width=3.5cm]{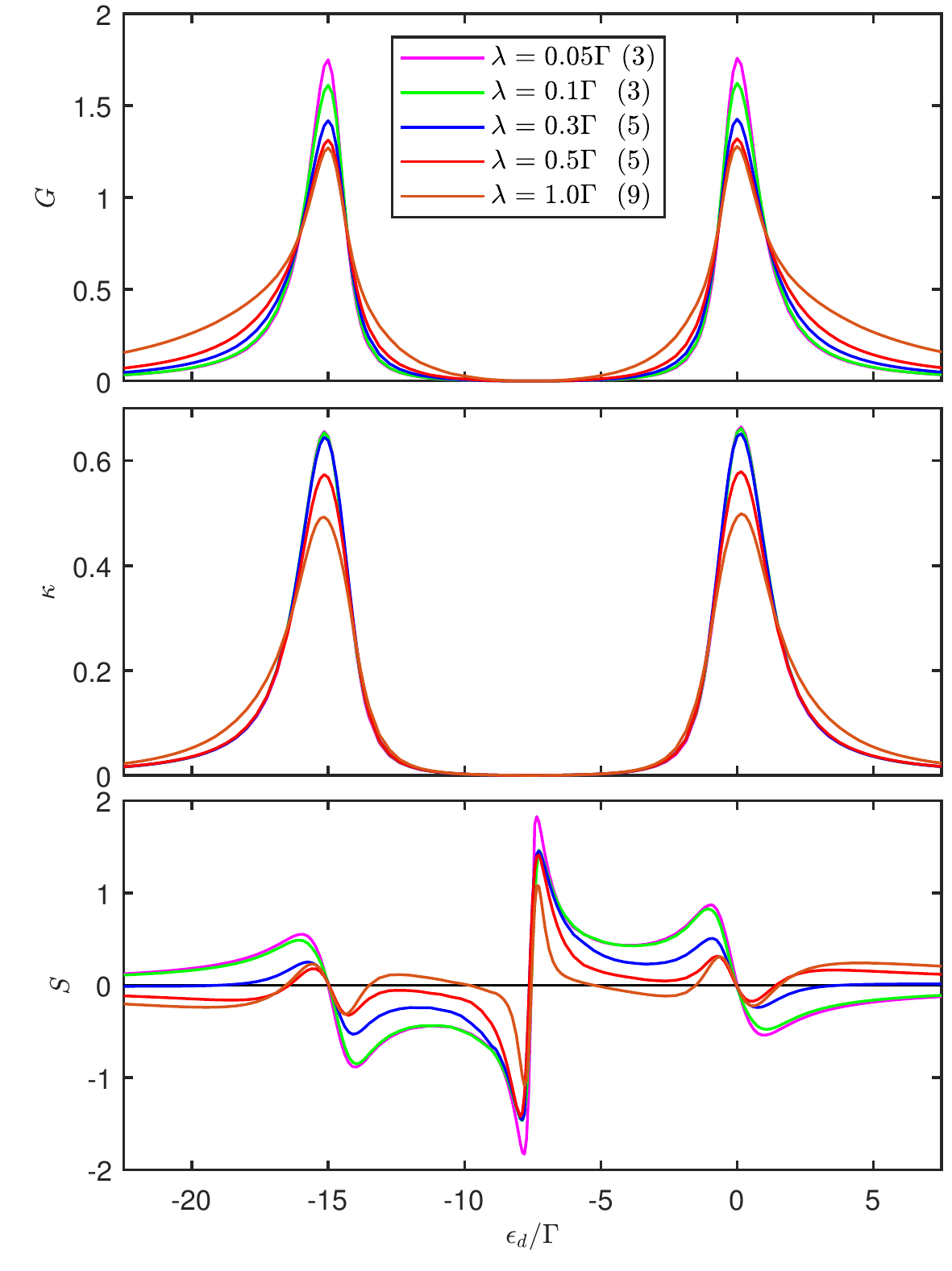}}
		\subfigure[$k_B T=0.15\Gamma$, $U=20\Gamma$]{\includegraphics[width=3.5cm]{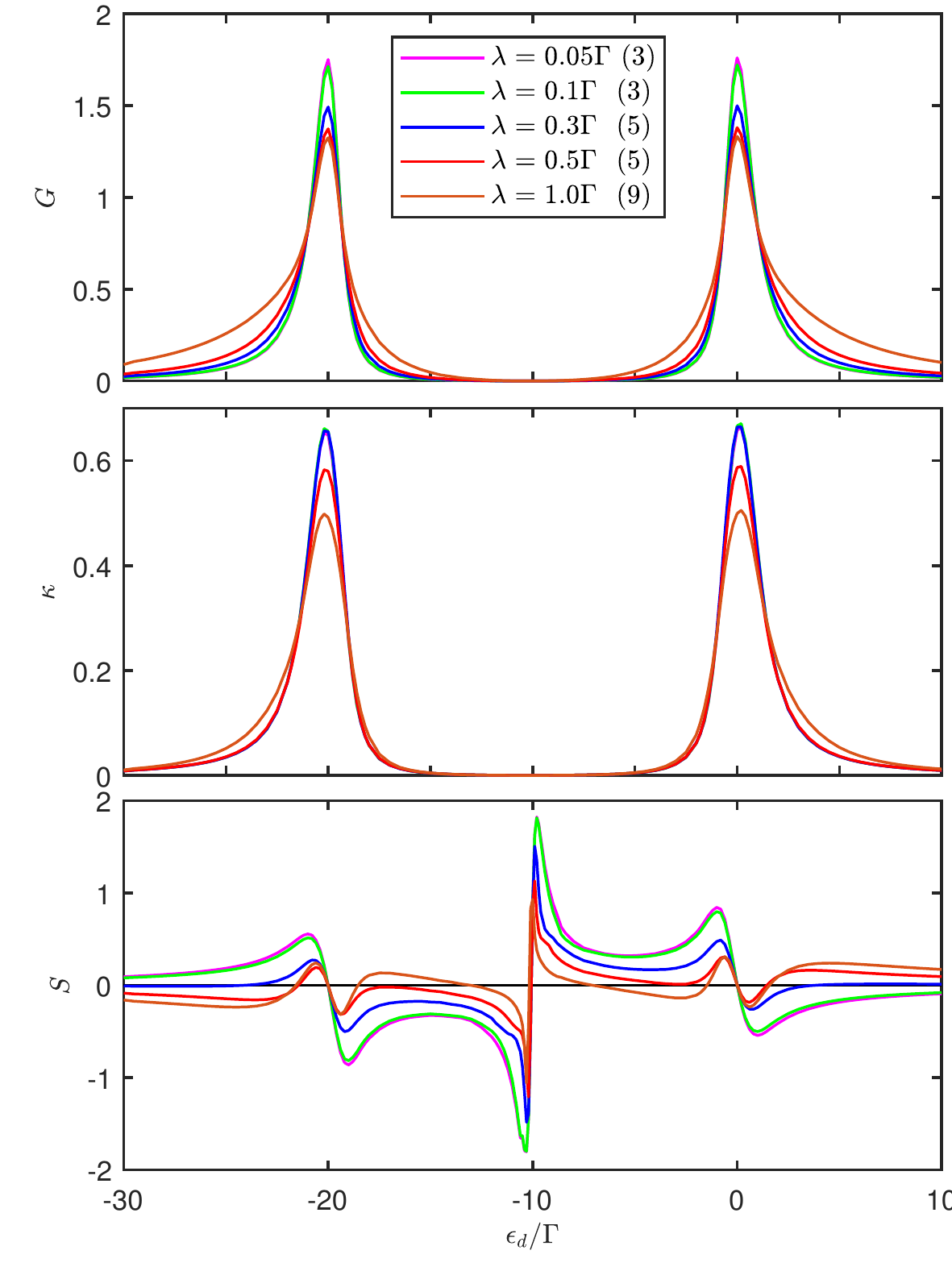}}
	\end{center}
	\caption{ 
		The electric conductance $G(e^2/h)$, the thermal conductance $\kappa(k_B/h)$ and the thermopower $S(k_B/h)$ for $\epsilon_M=0$ and various QD-MBS coupling $\lambda$, temperature $k_B T$ and  Coulomb interaction $U$. The number between parentheses next to the annotated values of $\lambda$ indicates the number of sign changes in the thermopower $S$.
	}
	\label{fig:thermo_charact_all}
\end{figure*}

Next, we consider the thermoelectric characteristics for $\epsilon_M=0$ and various QD-MBS coupling $\lambda$, temperature $k_B T$ and Coulomb interaction $U$ (see Fig. \ref{fig:thermo_charact_all}). 

Let us consider first the $G$, $\kappa$ and $S$ for various $\lambda$ with $kT=0.1\Gamma$, $U=10\Gamma$ [see Fig. \ref{fig:thermo_charact_all}(j)].
For large $\lambda(=0.5\Gamma,1.0\Gamma)$, the resonance characteristics of the electric conductance $G$ and thermal conductance $\kappa$ are not sensitive to the change of $\lambda$ and the sign of $S$ still changes 9 times.
Except for quantitative differences in $S$-graph, there exist little changes of the points $S=0$ according to the $\lambda$.
It is why the larger $\lambda$ is, the wider the width of MBS peak is and the larger the lean of that is.
However, for small $\lambda(=0.05\Gamma,0.1\Gamma)$, the sign of $S$ changes three times.
Since the integral Eq. (\ref{eq:def_I}) is accurately related to $E f'(E)$ rather than $f'(E)$, the $\textrm{DOS} (E)$ near $E = 0$, which $\abs{E} \ll k_B T$, makes only a small contribution to the sign of thermopower.
Therefore, the MBS peak with the very small width makes little contribution to the sign of $S$.
Since the width of the MBS peak is proportional to $\lambda^2$, the sign of thermopower for a very small $\lambda$ changes three times, as in $\lambda=0$.
For a very small $\lambda$, the resonance characteristics of $G$ and $\kappa$ are almost the same as for $\lambda=0$.
It is surprising that the sign of thermopower $S$ for a medium $\lambda(=0.3\Gamma)$ change 5 times.
As mentioned above, the widths and leans of the MBS peaks lying between the two Hubbard peaks [see Fig. \ref{fig:DOS_zero_why}(l),(o)] are less than those of the MBS peaks lying outside [see Fig. \ref{fig:DOS_zero_why}(c),(f),(l)], so for a medium $\lambda(=0.3\Gamma)$ former MBS peaks give a small contribution to the S-sign and do not change it.

It is very interesting to consider the influence of temperature $k_B T$ to the thermoelectric characteristics.
The higher temperature makes a little increasing of electric conductance $G$, because the resonant tunnelling is proportional to the width of $k_B T$.
The thermal conductance $\kappa$ becomes much larger than $G$, because there exist above effect and the charge carriers carry out the energy $k_B T$.
The sign change of $S$ according to the change of $k_B T$ is noticeable.
In the case of $k_B T=0.15\Gamma$ [Fig. \ref{fig:thermo_charact_all}(n)], for the $\lambda = 0.05\Gamma$, $0.1\Gamma$, $0.3\Gamma$, and $1.0\Gamma$, the changes of $S$-sign are the same as for $k_B T=0.1\Gamma$, but for the $\lambda=0.5\Gamma$, the $S$-sign changes 9 times.
It is because the higher the temperature, the larger the width of the thermal activation window $\Omega=k_B T$, so that the contribution of the MBS peak to the sign of thermopower becomes smaller and the contribution of the Hubbard peak increases (The DOS (\ref{eq:def_DOS}) is related to the temperature $k_B T$, however, the DOS resulted in our calculation is not actually sensitive to $k_B T$).
When the temperature goes down, the situation is reversed: 
In the case of $k_B T=0.07\Gamma$ [Fig. \ref{fig:thermo_charact_all}(g)] the number of $S$-sign changes for $\lambda=0.3\Gamma$ is 9, whereas in case of $k_B T=0.05\Gamma$ [Fig. \ref{fig:thermo_charact_all}(c)] the $S$-sign for $\lambda=0.1\Gamma$ changes 5 times.
It should receive emphasis that if the sign of $S$ changes 5 or 9 times, then $S$-graph starts from minus and lasts plus, but if it changes 3 times, $S$ starts with plus and end with minus (as in the case of $\lambda=0$).

The Coulomb correlation parameter $U$ does not qualitatively change the thermoelectric properties. 

\begin{figure}[t]
\begin{center}
  \includegraphics[width=8.0cm]{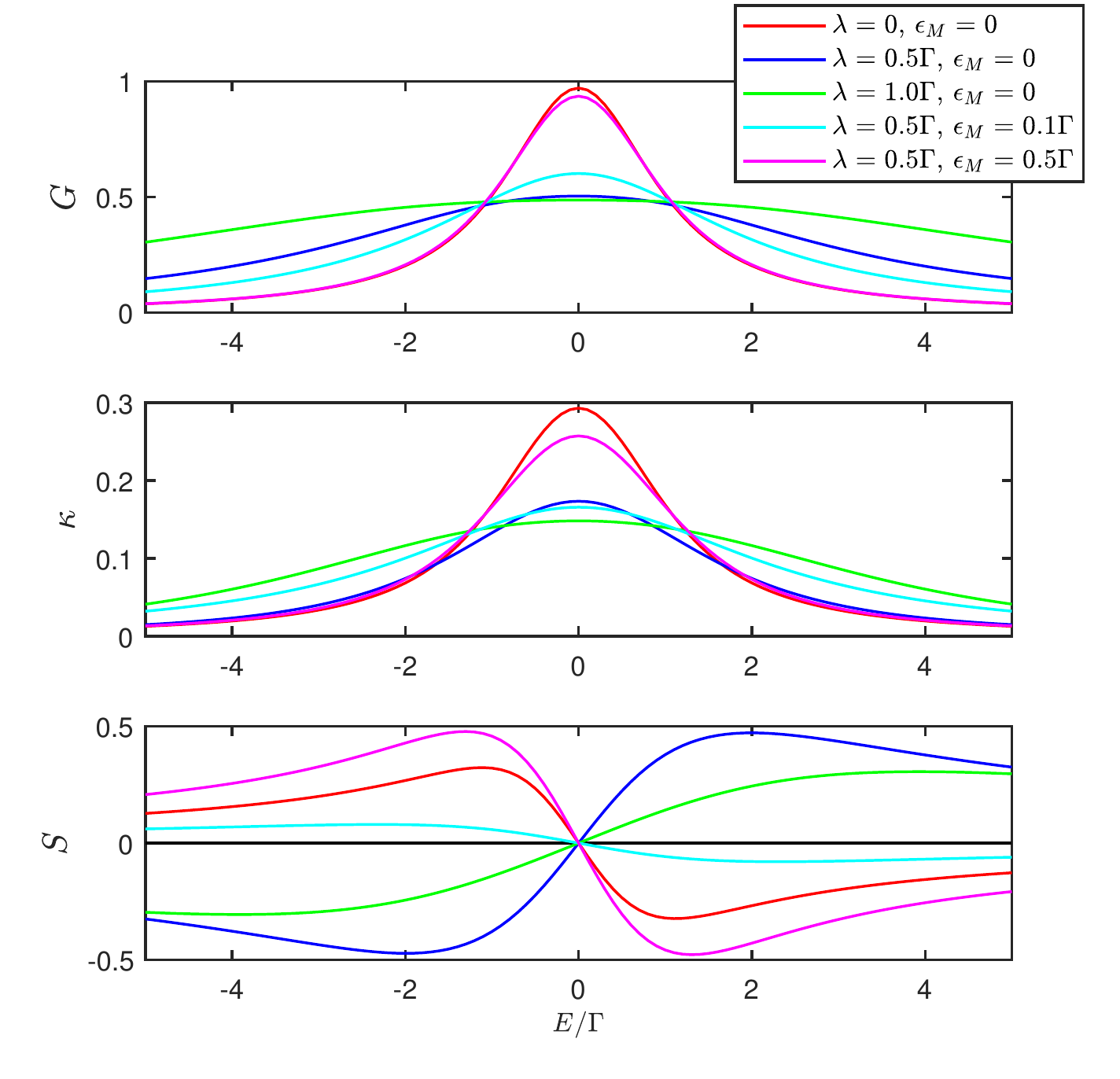}
\end{center}
	\caption{ 
		The electric conductance $G(e^2/h)$, the thermal conductance $\kappa(k_B/h)$ and the thermopower $S(k_B/h)$ as a function of $\epsilon_d$ for huge external magnetic field ($E_z=100\Gamma$). The other parameters are $U=10\Gamma$, $k_B T=0.1\Gamma$.
	}
	\label{fig:Zeeman_huge}
\end{figure}

Finally, we discuss the thermoelectric characteristics by supposing very huge external magnetic field ($E_z=100\Gamma$).
As shown in Fig. \ref{fig:Zeeman_huge}, for $\epsilon_M=0$ the electric conductance $G$ and thermal conductance $\kappa$ form one resonant peak and for $\lambda\neq0$ its maximum reduces half than one for $\lambda=0$.
The sign of thermopower changes once near the $\epsilon_d=0$, while the sign for $\lambda=0$ is opposite with one for $\lambda\neq0$.
For non-zero $\epsilon_M$, it shows no qualitative differences with resonant level model where $\lambda=0$.
In total, the previous result \cite{Lopez_PRB89} is remerged as it was.
It has turned out that under the huge external magnetic field one can regard it as appropriate that QD can be also considered as spinless QD.   

\section{Conclusions}\label{sec:conclusions}

In this paper we have studied on the thermoelectric transport through single-level QD side-coupled to MBS and presented the influence of MBS to the characteristics of thermoelectric transport through QD.
Under not so large magnetic field Coulomb interaction in QD is considered, which agrees with the recent experiment \cite{Deng_SCI354}. 
We calculate the QD Green function represented by 4-component Nambu spinor formalism by using the EOM method in the framework of nonequilibrium Green function technique. 
To focus on pure effect of MBS, we consider only the relatively high temperature region ($T\gg T_K$), where Kondo effect does not appear, so use the Hartree-Fock approximation. 

The electric and thermal conductance and thermopower as a function of gate voltage (i.e. QD level) are completely different whether $\epsilon_M$ is zero or not.
For non-zero $\epsilon_M$, all characteristics are nearly the same with in the normal case without MBS.
However, for $\epsilon_M=0$, the height of the resonant peak in electric and thermal conductance is reduced by about 3/4 than the one without MBS.
The behaviour of thermopower $S$ is very interesting.
In the case of normal QD without MBS, the sign of thermopower changes three times, however, in the case of QD side-coupled to ideal and isolated MBS ($\epsilon_M=0$), the sign of thermopower changes 9, 5 and 3 times for a large, medium and small QD-MBS coupling $\lambda$, respectively.
Such complicated behaviour of the sign in thermopower is why the MBS peak near $E=0$ leans to the left or right due to the shifting effect by interacting with two QD effective levels.
Such behaviour of $S$ is remaining as ever for different Coulomb correlation.
The number of the sign changes of $S$ for a given $\lambda$ is not fixed but varies with temperature.
As the temperature increases, the number of the $S$-sign changes corresponding to the $\lambda$ giving the small width of MBS peak that does not contribute to the $S$-sign change increases.
If is important that when the sign of $S$ changes 5 or 9 times, $S$-graph starts from minus and lasts plus. For a very small $\lambda$, all characteristics are similar to those in the normal case without MBS.
Finally we have showed that for huge magnetic field, the thermoelectric characteristics are similar with spinless QD's. 

It is regarded that the fact that the sign of the thermopower in QD strongly side-coupled to ideally isolated MBS changes 9 or 5 times and the electrical and thermal conductance are reduced by 3/4 can also be used for detecting of the signature of MBS.
Maybe, to measure the change of the sign of $S$ is relatively easier and does not require the higher accuracy than to measure the exact numerical values.
Furthermore, since the above characteristics are remaining as ever when the coupling between two MBSs is very small, it has actual possibilities when the nanowire is long enough and pure without any defects. 

The change of sign in thermopower is related to behaviour of DOS at $E=0$.
At very low temperature Kondo peak appears near $E=0$ and it should interact with MBS to make change of the sign in thermopower more complicatedly.
It will be possible to study the properties above by using the higher order of approximations beyond Hartree-Fock approximation. 

\section{Authors contribution statement}
All authors contributed equally to the paper.

\begin{acknowledgement}
K. H. Jong wishes to thank Prof. A. N. Nersesyan and M. N. Kiselev for helpful advices.
This work is supported by the National Program on Key Science Research of Democratic People's Republic of Korea (Grant No. 18-1-3).
\end{acknowledgement}

\end{document}